\documentclass[twoside]{article}
\usepackage{amsmath,amsfonts,amssymb,mathrsfs,bm}
\usepackage{qic,epsfig}
\usepackage{tikz}
\usepackage{cite}
\usepackage{color}
\usepackage{booktabs}
\usepackage{cases}
\usepackage[mathscr]{euscript}
\usepackage{algorithm}
\usepackage{algpseudocode}
\usepackage{graphics}
\usepackage{enumitem}
\usepackage{xcolor}

\renewcommand{\theequation}{\arabic{section}.\arabic{equation}}

\newtheorem{prop}{Proposition}
\newtheorem{thm}{Theorem}
\newtheorem{cor}{Corollary}
\newtheorem{lem}{Lemma}

\renewenvironment{proof}{\par\noindent{\bf Proof.}}{$\quad\Box$\par}
\newcommand{\ket}[1]{| #1 \rangle}
\newcommand{\bra}[1]{\langle #1 |}

\begin{document}

 \date{}

\setlength{\textheight}{8.0truein}    

\runninghead{Exact Homomorphic Encryption} { }

\normalsize\textlineskip \thispagestyle{empty}
\setcounter{page}{1}

\vspace*{0.88truein}

\alphfootnote

\fpage{1}

\centerline{\bf Exact Homomorphic Encryption}
\vspace*{0.035truein} \centerline{\footnotesize Zheng-Yao
Su\footnote{Email: zsu@narlabs.org.tw}\hspace{.15cm}and Ming-Chung
Tsai} \centerline{\footnotesize\it National Center for
 High-Performance
 Computing,}
 \centerline{\footnotesize\it National Applied Research Laboratories,
 Taiwan, R.O.C.}

\vspace*{0.21truein}

\abstracts{
 Inspired by the concept of fault tolerance quantum computation,
 this article proposes a framework dubbed {\em Exact Homomorphic Encryption},
 {\em EHE},
 enabling exact computations on encrypted data without the need for pre-decryption.
 The introduction of {\em quantum gates} is a critical step for constructing the message encryption and the computation encryption within the framework.
 Of significance is that both encryptions are respectively accomplished in a multivariate polynomial set generated by quantum gates.
 Two fundamental traits of quantum gates,
 the {\em invertibility} and the {\em noncommutativity},
 establish the success of EHE.
 The encrypted computation is {\em exact} because its encryption transformation is conducted with invertible gates.
 In the same vein, decryptions for both an encrypted message and encrypted computation are {\em exact}.
 The second trait of {\em noncommutativity} among applied quantum gates brings forth the security for the two encryptions.
 Toward the message encryption, a plaintext is encoded into a ciphertext via a polynomial set generated
 by a product of {\em noncommuting} gates {\em randomly} chosen.
 In the computation encryption, a desired operation is encoded into an {\em encrypted polynomial set} generated by another product of noncommuting gates.
 The encrypted computation is then the evaluation of the encrypted polynomial set on the ciphertext and is referred to as the {\em cryptovaluation}.
 On the basis of the difficulty of retrieving an action from its encrypted polynomial set,
 the cryptovaluation is considered a {\em blind computation}.
 Attributed to the fact that each quantum gate adopted is a mapping of {\em dimension-one preserving},
 EHE is not only attainable on quantum computers,
 but also straightforwardly realizable on traditional computing environments.
 Surpassing the standard security $2^{128}$ of {\em quantum resilience},
 both the encryptions further reach a security greater than the suggested threshold $2^{1024}$
 and are characterized as {\em hyper quantum-resilient}.
 Thanks to the two essential traits of quantum gates,
 this framework can be regarded as the initial tangible manifestation of the concept {\em noncommutative cryptography}.
 The software required to implement EHE over diverse functions
 is available on CPU and GPU computing environments.  }{}{}

 \vspace{3pt} \vspace*{1pt}\textlineskip

 \section{Introduction}
\label{secintro}
  \renewcommand{\theequation}{\arabic{section}.\arabic{equation}}
\setcounter{equation}{0} \noindent
 Homomorphic Encryption (HE), hailed as ``the Holy Grail of encryption",
 is deemed the most important subject in cryptography~\cite{2015-HE-HolyGrail-Wu,2018-AAAC-HE-Review,2018-HE-Review-RL}.
 It permits users to compute on encrypted messages without prior decryption,
 thus providing a high level of security for the data processing.
 First introduced by Rivest et al. in 1978,
 this concept aims to arrive at Fully Homomorphic Encryption (FHE)~\cite{HE-1978}.
 Over the next 30 years, FHE was unsuccessful until Gentry's work in 2009~\cite{HE-Gentry-2009}.
 His dissertation puts forward a technique called ``bootstrapping" operated on ideal lattices,
 transforming the somewhat HE of a function into FHE~\cite{HE-Gentry-2009}.
 The accumulation of noise poses a hindrance to execute this technique.
 The predicament is especially pronounced by dint of the exponential growth of noise with the number of multiplications~\cite{2018-AAAC-HE-Review,2022-HE-LWE-DK}.
 Subsequent to Gentry's work,
 a plethora of studies on lattice-based HE have emerged for refining the bootstrapping blueprint.
 The objective involves noise controls~\cite{HE-BGV-2011,HE-GHS-2012,2018-AAAC-HE-Review,HE-RLWE-2017-Kerskin,2019-HE-standard}
 or managements~\cite{2022-HE-LWE-DK,2018-AAAC-HE-Review,HE-RLWE-2017-Kerskin,2014-LevelHE-BGV,2017-HE-CKKS,2019-HE-standard}
 during a homomorphic computation.
 However,
 a certain number of obstacles occur in these intentions.
 Being proportional to the square of security parameter and increasing nonlinearly with the lattice dimension,
 the size of ciphertext imposes restrains on the scalability of encryption~\cite{2017-HE-CKKS,2019-HE-standard,2022-HE-LWE-DK}
 and other vulnerabilities that compromise the security~\cite{2019-Lattice-secure-compare-Bernstein,2018-HE-Review-RL}.
 Because each ciphertext is corrupted by errors and the interference rises quadratically per calculation of cryptograms,
 there educe only approximate encrypted computations ensuing from noise reductions~\cite{2017-HE-CKKS,2019-HE-standard}.
 Governing the noise escalation narrows choices of operations and limits the computation depth~\cite{HE-BGV-2011,2014-LevelHE-BGV}.
 Types of enciphered functions are unmasked and identifiable throughout the whole trial,
 causing extra security threats.
 The decryption undergoes a failure probability that is heightened nonlinearly with the input size of encryption,
 which engenders offensives that diminish the security~\cite{2009-LatticeCrypt-Regev,2016-decode-fail-code-based-GJS-858,2018-decoding-fail-AVV-1089}.
 Briefly, the deficiencies exposed constrict the scales of encrypted computations reachable in current HE.

 Quantum computing has garnered much attention recently inasmuch
 as its momentous influence not only on data processing~\cite{ShorAlg,GroverAlg,2001-Shor-EXP-Chuang,1998-Grover-EXP-Chuang},
 but also on information protection.
 An intriguing field of study in relation to the security hazard is Quantum Public-Key Encryption (QPKE).
 The core approach entails the production of one-way functions
 to generate a quantum state that plays the role of a public key for encrypting message~\cite{2005-QPKE-Kawachi,2023-QPKE-BMW,2023-QPKE-GOV,2023-QPKE-Kitagawa}.
 Other than the debates over PKE,
 such as the public-key certificate~\cite{2005-QPKE-Kawachi,2023-QPKE-Kitagawa},
 noisy decryption~\cite{2023-QPKE-GOV,2023-QPKE-Kitagawa} and security proofs~\cite{2023-QPKE-BMW,2023-QPKE-GOV},
 QPKE is impeded mainly by necessitating sizable quantum operations,
 which falls into the hurdle of scaling up quantum computers~\cite{1998-FTQC-JP,2000-FTQC-BL,2022-FTQC-Gotts}.
 Quantum Homomorphic Encryption (QHE) is another research area that has become increasingly appealing
 to safeguard data manipulation~\cite{2017-TransGateHE-NS,2015-QHE-BJ,2017-QHE-Ouyang,2018-QHE-EXP-Tham,2018-QHE-LWE-Brak,2021-QHE-LWE-CDM,2021-QHE-Code-LY}.
 Typically, an encrypted computation is exercised with a fault-tolerant Clifford$+$T circuit~\cite{2015-QHE-BJ,2017-TransGateHE-NS,2017-QHE-Ouyang,2018-QHE-EXP-Tham}.
 Formed in transversal gates on a few quantum codes restrictively,
 the Clifford$+$T fault tolerance may not lend itself well to large-scale quantum computations~\cite{2017-TransGateHE-NS}.
 Explicitly,
 physical qubits outnumber logical qubits by at least several hundred times,
 refuting the accessibility of QHE.
 An alternative rephrases a present HE to its quantum version~\cite{2018-QHE-LWE-Brak,2021-QHE-LWE-CDM,2021-QHE-Code-LY}.
 Aside from receiving the demerits of HE schemes aforesaid,
 the method in view consumes numerous qubits and then encounters the scalability barrier of quantum computers.

 A serial of episodes elucidates a structure called the {\em Quotient Algebra Partition}, {\em QAP},
 universally existing in finite-dimensional unitary Lie algebras~\cite{OriginQAPSu,QAPSu0,QAPSu1,QAPSuTsai1,QAPSuTsai2}.
 Given this structure inherited by every stabilizer code,
 a general methodology of Fault Tolerance Quantum Computation in QAP~\cite{SuTsaiQAPFT},
 abbreviated as QAPFTQC,
 elicits an algorithmic procedure achieving the acquirement that every action in every error-correcting code is fault tolerant.
 That is, on a code, a quantum state is encoded into a codeword
 and a target operation is encoded into an encrypted action, called the fault tolerant encode.
 A fault tolerance quantum computation is thence derived by applying this encode on the codeword.
 Stemming from the concept of QAPFTQC, the framework {\em Exact Homomorphic Encryption}, {\em EHE},
 is proposed to admit computations on encrypted data.
 The message encryption and the computation encryption of EHE
 are thought of as analogous to the cryptograph of a quantum state and the fault-tolerant counterpart of a computation in QAPFTQC.

 Introducing {\em quantum gates},
 each of which is a {\em unitary transformation} in a Hilbert space and thus {\em invertible},
 to the construction of EHE is a critical step.
 An important leap forward is that invertible gates take the place of non-invertible logic operations
 leveraged in finite computations~\cite{universalboolean-0th,UniversalBoolean-1st}.
 Not only acting on quantum states as commonly known, in EHE,
 quantum gates are also applied on {\em variables} to generate {\em polynomials}.
 This ingenuity induces the fruition of encrypting the message and the computation respectively using a {\em multivariate polynomial set}.
 In the message encryption,
 a polynomial set, generated by an {\em encryption mapping} consisting of quantum gates,
 serves as a {\em public key} that encodes a message into a ciphertext.
 Specifically, the ciphertext is the {\em evaluation} of the polynomial set on the message,
 {\em i.e.}, calculating the polynomial set with the message as an input.
 The cryptographic primitive of a message is adaptable to the computation encryption.
 A desired operation is cryptified into an encrypted action through an {\em encryption transformation} composed of quantum gates.
 The {\em circuit} of the encrypted action is represented in an {\em encrypted-computation polynomial set},
 or abbreviated as an {\em encrypted polynomial set}.
 Evaluating the encrypted polynomial set on the ciphertext yields the encrypted computation,
 termed as {\em cryptovaluation}.
 The triumph of the encryptions of message and computation is ascribable to the
 a {\em duality relation} that portrays the equivalence of the {\em polynomial evaluation} and {\em state computation}.

 The trait of {\em invertibility} of quantum gates assures the {\em exactness} for both encryptions of message and computation.
 In contrast to approximated homomorphic computations
 of moderate size of current HE schemes~\cite{2018-AAAC-HE-Review,2018-HE-Review-RL,HE-RLWE-2017-Kerskin,HE-BGV-2011,HE-GHS-2012,2019-HE-standard},
 evaluating the encrypted polynomial set generated by a product of {\em invertible} gates actualizes an {\em exact} cryptovaluation of extensive scale.
 On a similar note, the decryption is {\em accurate} exploiting the private key comprising invertible gates,
 instead of the erroneous deciphering for cryptosystems in existence~\cite{2009-LatticeCrypt-Regev,2018-decoding-fail-AVV-1089,HE-GHS-2012,2019-HE-standard}.
 Every activated gate enjoys the characteristic of {\em dimension-one preserving} that transforms a basis quantum state into another.
 Thus, each computation of EHE needs not the ample memory of simulating a quantum computation.
 Different from the floundering of quantum cryptographic attempts,
 the two encryptions are practicable on traditional computing environments with no reliance on quantum computers.
 EHE possesses a high level of security arising from the trait of {\em noncommutativity} of quantum gates.
 Towards attacking either of the two encryptions,
 a {\em combinatorially} high complexity is demanded on the retrieval of a {\em circuit} of {\em noncommuting} gates.
 This process is shown to be harder than solving approved intractable problems.
 Coalescing with the concealment of types of encoded functions,
 the cryptovaluation is further perceived as a {\em blind computation},
 a distinction absent from an existent HE.
 Alongside exceeding the {\em quantum-resilient} standard of security $2^{128}$,
 the two encryptions forwardly surpass the threshold $2^{1024}$ of the propounded {\em hyper quantum resilience}.
 In this light, the framework EHE is the unprecedented incarnation of ``noncommutative cryptography",
 a notion previously explored abstractly without concrete instantiations~\cite{noncommute-2011-book,noncommute-2020}.
 The software devoted to the EHE framework is readily deployable~\cite{EHE-software-test-web}.

 This article is organized as follows.
 Section~\ref{secArithmQCircuit} familiarizes readers with the most basic quantum gates in the EHE framework,
 manifesting the evaluation duality of a polynomial set and the associated state.
 Next, in Section~\ref{secMessageEncrypt},
 the mechanism of a highly secure message encryption is expounded.
 Section~\ref{secHECompute} delineates the enciphering of computations at an advanced level of security.
 The experimental outcomes of the EHE software on CPU and GPU environments are provided in Section~\ref{secExps},
 including encrypted computations of addition, subtraction, multiplication, division, string comparison, sum of squares and monomial powers~\cite{EHE-software-test-web}.
 The concluding section summarizes the attainments of research efforts and emphasizes the visions of EHE in the future.

 \section{Elementary Gate}\label{secArithmQCircuit}
  \renewcommand{\theequation}{\arabic{section}.\arabic{equation}}
\setcounter{equation}{0}\noindent
 In structuring the framework EHE, the introduction of quantum gates is a pivotal step.
 Prominently,
 non-invertible logic operations, wielded in finite computations~\cite{universalboolean-0th,UniversalBoolean-1st},
 are replaced with quantum gates.
 This section will disclose
 the most basic units of quantum gates that fulfill the {\em computational universality}.
 Apart from operating on quantum states,
 each employed gate also acts on {\em variables} to generate {\em multivariate polynomials}.
 This perceptive notion leads out of the debut of the {\em duality} that conjoins the {\em polynomial evaluation} and {\em state computation},
 which is essential for the encryptions of message and computation.

 To compile the computation model of Turing machine,
 classical logic operations are basic building entities of writing every finite function
 as a Boolean circuit of finite size~\cite{2009-BooleanCircuitUniversal-AB,universalboolean-0th}.
 Among logic operations,
 there exists a {\em universal set of gates} that is functionally complete in fabricating each finite Boolean circuit~\cite{2009-BooleanCircuitUniversal-AB,UniversalBoolean-1st}.
    \vspace{6pt}
 \begin{prop}\label{propTuringCompleteBoolean}
 Every finitely computable function is achievable by a
 finite-size circuit consisting of logic operations of negation, AND, and OR.
 \end{prop}
 \vspace{6pt}
 These three types of operations form a {\em universal set} of Boolean arithmetic~\cite{2009-BooleanCircuitUniversal-AB,UniversalBoolean-1st}.

 It comes to spelling each basic Boolean logic in quantum gates.
 Given an encapsulation of Pauli matrices
 ${\cal S}^{\epsilon_i}_{a_i}=\ket{0}\bra{a_i}+(-1)^{\epsilon_i}\ket{1}\bra{1+a_i}$~\cite{QAPSu0,QAPSu1,SuTsaiQAPFT},
 the {\em $s$-representation} casts an $n$-qubit spinor into the expression
 ${\cal S}^{\zeta}_{\alpha}={\cal S}^{\epsilon_1\epsilon_2\cdots\epsilon_n}_{a_1a_2\cdots a_n}
 ={\cal S}^{\epsilon_1}_{a_1}\otimes{\cal S}^{\epsilon_2}_{a_2}\otimes\cdots\otimes{\cal S}^{\epsilon_n}_{a_n}$,
 where $\alpha$ is a {\em bit string} and $\zeta$ is a {\em phase string},
 $a_i,\epsilon_i\in{Z_2}$ and $i=1,2,\cdots,n$.
 In this formulation, the spinor ${\cal S}^{\zeta}_{\alpha}$ maps an $n$-qubit basis state $\ket{\beta}$
 to another ${\cal S}^{\zeta}_{\alpha}\ket{\beta}=(-1)^{\zeta\cdot(\beta+\alpha)}\ket{\beta+\alpha}$, $\beta\in{Z^n_2}$;
 here ``$\cdot$" denotes the {\em inner product} and ``$+$" the {\em bitwise addition}.
 Literally, the negation is a 1-qubit spinor.
 \vspace{6pt}
 \begin{lem}\label{lemNOTgateQuantum}
 Every negation is a single-qubit spinor ${\cal S}^0_1$.
 \end{lem}
 \vspace{2pt}
 \begin{proof}
 The proof is clear that ${\cal S}^0_1\ket{a}=\ket{a+1}$
 for a one-qubit spinor ${\cal S}^0_1$ acting on a basis state $\ket{a}$, $a\in{Z_2}$~\cite{QAPSu0,QAPSu1}.
 \end{proof}
 \vspace{6pt}

 Ancilla qubits are necessary to the invertible compositions of AND and OR.
   \vspace{6pt}
 \begin{lem}\label{lemANDgateQuantum}
 Every AND operation can be expressed as a Toffoli gate plus an ancilla qubit.
 \end{lem}
 \vspace{2pt}
 \begin{proof}
 With the additional qubit $\ket{0}$,
 an AND for the two inputs $a_1$ and $a_2\in{Z_2}$
 is minimally written in $T^{12}_3\ket{a_1,a_2,0}=\ket{a_1,a_2,a_1\cdot a_2}$.
 Here, $T^{12}_3$ is a Toffoli gate conditioning the 3rd qubit by the first two~\cite{Barenco1995}
 and $\ket{a_1,a_2,0}$ is a 3-qubit basis state.
 The bitwise multiplication $a_1\cdot a_2=a_1\wedge a_2$ is exactly the AND logic
 adhering to $a_1\wedge a_2=1$ if $a_1=a_2=1$ and $a_1\wedge a_2=0$ otherwise.
 \end{proof}
 \vspace{6pt}
 \vspace{6pt}
 \begin{lem}\label{lemORgateQuantum}
 Every OR operation can be represented in a composition of three negations and a single Toffli gate
 with an ancilla qubit.
 \end{lem}
 \vspace{2pt}
 \begin{proof}
 This lemma is straightforwardly derived,
 \begin{align}\label{eqOROP-Quantum}
 T^{12}_3{\cal S}^{000}_{111}\ket{a_1,a_2,0}=T^{12}_3\ket{a_1+1,a_2+1,1}
 =\ket{a_1+1,a_2+1,1+(a_1+1)\cdot(a_2+1)},
 \end{align}
 where ${\cal S}^{000}_{111}$ is a tensor product of three
 negations,
 $T^{12}_3$ a Toffoli gate and $\ket{a_1,a_2,0}$ a basis state,
 $a_1$ and $a_2\in{Z_2}$.
 The value on the accessory qubit $1+(a_1+1)\cdot(a_2+1)=a_1\vee a_2$ is the outcome of an OR logic,
 {\em i.e.}, $a_1\vee a_2=0$ if $a_1=a_2=0$ and $a_1\vee a_2=1$ otherwise.
 \end{proof}
 \vspace{6pt}

\begin{figure}[!ht]
 \centering 
 \scalebox{0.5} 
 {\includegraphics{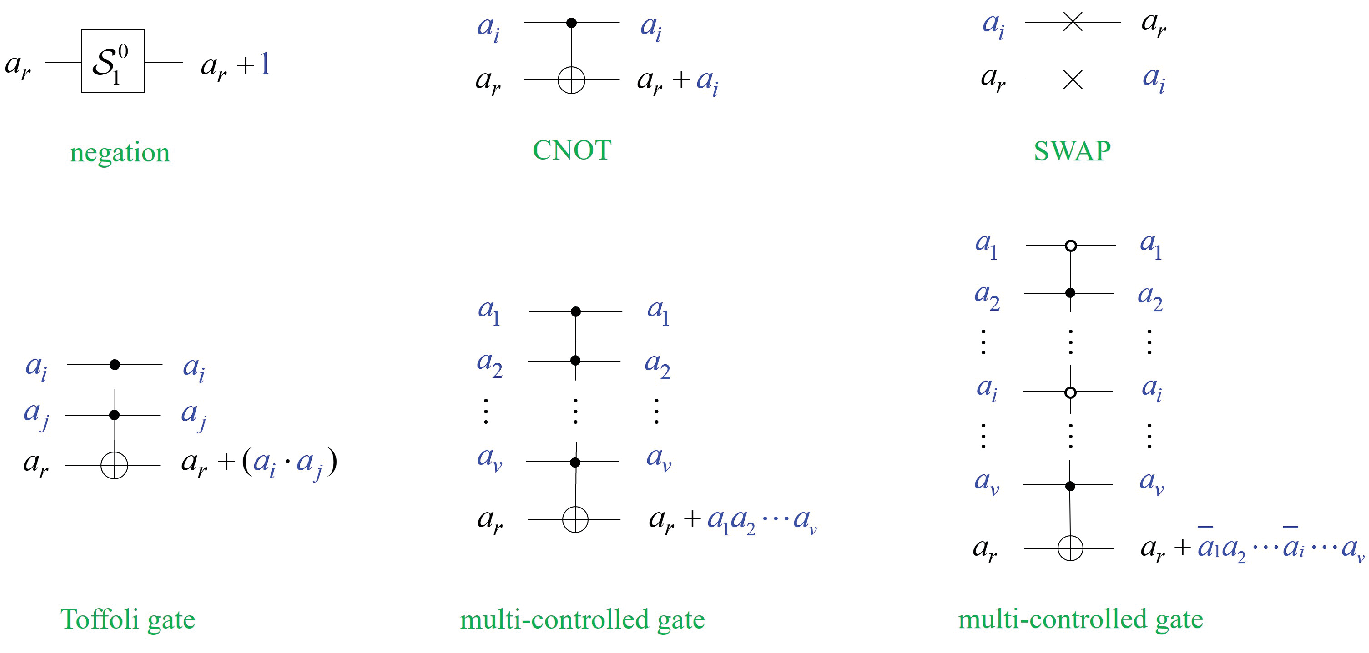}}
 \vspace{5pt} 
 \fcaption{The set of elementary gates,
 where ``$\oplus$" denotes the target bit controlled by a single
 bit or multiple bits,
 each control bit is denoted as the black dot ``$\bullet$" if it is in the value $1$
 or as the white dot ``$\circ$" if it is in $0$,
 and $\bar{a}_i=a_i+1$ is the flipping of $a_i\in{Z_2}$,
 $i,j,r,v\in[k]$.~\label{Fig-ElementaryGate}}
\end{figure}

 By precedent assertions,
 every finitely computable function is implementable in negations and Toffoli gates with auxiliary qubits.
 Incorporating the other two kinds of quantum gates into EHE is convenient for the raise of performance.
 The first choice is the {\em Controlled-Not (CNOT)} gate,
 a basic quantum operation that can be formed in negations and Toffoli gates plus ancilla qubits~\cite{Barenco1995};
 a SWAP is a composition of three CNOTs.
 The {\em multi-controlled gate}, targeting a qubit by multiple control bits, is another member to be included.
 Each $k$-qubit multi-controlled gate is expressible exploiting $O(k)$ Toffoli
 gates and an appropriate number of extra qubits~\cite{Barenco1995}.
 The basic units of quantum gates are thus bestowed.
 \vspace{6pt}
 \begin{prop}\label{propelementaryGates}
 The set of elementary gates consists of four kinds of operations, the negation, the CNOT, the Toffoli, and the multi-controlled gates.
 \end{prop}
 \vspace{6pt}
 Every elementary gate is a transformation of {\em dimension-one preserving} that maps a basis quantum state into another,
 referring to Fig.~\ref{Fig-ElementaryGate} for the diagrammatic exemplification.
 Since AND and OR can be rephrased in Toffoli gates attended with ancilla qubits,
 this set vouches for the {\em computational universality}.
    \vspace{0pt}
 \vspace{0pt}

 Simulating quantum computations on classical environments is presently challenging
 imputed to the hindrance of expensive memory cost that increases exponentially with the number of qubits
 required in realizing full states~\cite{2020-QC-simulate-memoryIssue,2018-QC-simulate-64Q}.
 Classical architectures available can emulate systems sizing up to $64$ qubits at the most~\cite{2018-QC-simulate-64Q}.
 The embodiment of quantum attempts of encryption~\cite{2005-QPKE-Kawachi,2023-QPKE-BMW,2023-QPKE-GOV,2023-QPKE-Kitagawa,2015-QHE-BJ,2017-QHE-Ouyang,2018-QHE-EXP-Tham,2018-QHE-LWE-Brak,2021-QHE-LWE-CDM,2021-QHE-Code-LY}
 on classical computers is hence limited by daunting resources demanded.
 The framework EHE enlists only the set of elementary gates,
 each of which is dimension-one preserving.
 This showcases that EHE avoids the burdensome memory spent to the simulation of a full state and is reachable on CPU and GPU,
 obviating the necessity of quantum computers.
 The trial data of computing enciphered {\em elementary functions} will be rendered in Section~\ref{secExps}.
 These functions encompass the addition, subtraction, multiplication, division, string comparison, sum of squares and monomial powers,
 referring to~\cite{QCArithmetic-1st,QCArithmetic-2nd,QCArithmetic-3rd,QCAddition-Ripple,QCAddition-Taka,QCMulti-Karatsuba}
 for optional circuits of calculations and to~\cite{EHE-software-test-web} for software tests of encrypted computations.

 As will be illustrated in upcoming sections,
 the framework EHE harnesses {\em multivariate polynomial sets} for encrypting message and computation.
 The construction is organized upon the concept that every elementary gate,
 alongside acting on quantum states conventionally,
 is also applied on {\em variables} to generate {\em polynomials} over the {\em binary field} $Z_2$.
 Ahead of delving into specifics, basic notations are furnished beforehand.
 A multivariate polynomial of $k$ variables $f(\bm{x})=\sum_{\tau\in{Z^k_2}}c_{\tau}\bm{x}^{\tau}$ is a linear combination of {\em monomials} $\bm{x}^{\tau}$
 of degrees $\leq k$ with coefficients $c_{\tau}\in Z_2$~\cite{1991-book-Boolean-Artin}.
 Each monomial $\bm{x}^{\tau}=x^{\sigma_1}_1x^{\sigma_2}_2\cdots x^{\sigma_k}_k$ is a product of powers $x^{\sigma_r}_r$
 of $k$ variables $x_r\in Z_2$,
 $\tau=\sigma_1\sigma_2\cdots\sigma_r\cdots\sigma_k\in{Z^k_2}$ and $r\in[k]$.
 Hereafter, the symbol $[k]$ denotes the set of positive integers from $1$ to $k$.
 To enunciate the notion of generating polynomials by applying elementary gates on variables,
 let the deliberation start with the seed equation.
 \vspace{6pt}
 \begin{prop}\label{Prop-defnGateonVar}
 Denoted as $\Lambda^{\theta}_{r}$,
 an elementary gate of $k$ qubits transforms a single variable $x_s\in{Z_2}$ into
 \begin{align}\label{eqcontrol-onMonomial}
 \Lambda^{\theta}_{r}\Vdash \hspace{1pt}x_s=x_s+\delta_{rs}\bm{x}^{\theta},
 \end{align}
 $r$ and $s\in[k]$,
 where $\bm{x}^{\theta}=x^{\epsilon_1}_1x^{\epsilon_2}_2\cdots x^{\epsilon_k}_k$ is a monomial of $k$ variables,
 the subscript $r$ signifies the $r$-th qubit as the target bit of $\Lambda^{\theta}_{r}$,
 and the nonzero entities of the $k$-bit binary string $\theta=\epsilon_1\epsilon_2\cdots\epsilon_k\in{Z^k_2}$
 evinces the positions of qubits that serve as control bits.
 \end{prop}
 \vspace{6pt}
 The mapping of Eq.~\ref{eqcontrol-onMonomial} {\em de facto} unveils the {\em polynomial representation} of elementary gates.
 Applied by this mapping,
 the variable $x_s$ receives a shift of the product $\bm{x}^{\theta}$
 if the $s$-th qubit is identical to the target bit, or remains intact otherwise.
 In practical maneuvers, elementary gates operate on variables of monomials.
 The gate $\Lambda^{\theta}_{r}$ is said to be of {\em rank t} if $\theta$ contains a number $t$ of nonzero bits.
 That is, a negation is of rank zero,
 a CNOT rank one, a Toffoli rank two,
 and a multi-controlled gate is of rank $t\geq 3$.
 Notice that every elementary gate defined in Proposition~\ref{Prop-defnGateonVar} is unitary and involutory.

 The most general definition of an elementary gate of $k$ variables over $Z_2$ may be written as
 \begin{align}\label{eqmostGenGateFormula}
 \Lambda^{\theta,\zeta}_r\Vdash \hspace{1pt} x_s=x_s+\delta_{rs}\bar{\bm{x}}^{\theta}_{\zeta},
 \end{align}
 where $r,s\in[k]$,
 $\theta=\epsilon_1\epsilon_2\cdots\epsilon_k$
 and
 $\zeta=\varsigma_1\varsigma_2\cdots\varsigma_k\in{Z^k_2}$,
 and
 $\bar{\bm{x}}^{\theta}_{\zeta}=\prod^k_{i=1}(x_i+\varsigma_i)^{\epsilon_i}$.
 The $i$-th bit of this gate is a control bit of ``black dot" if $\varsigma_i=0$ and of ``white dot" if $\varsigma_i=1$,
 {\em cf.} Fig.~\ref{Fig-ElementaryGate}.
 The white-dot bit is acquirable by sandwiching the black-dot with negations at the same position.
 As a result, it suffices to capitalize on the formulation of Eq.~\ref{eqcontrol-onMonomial},
 involving only control bits of black dot, in the ensuing exposition.

 Every elementary gate is applicable to quantum states as well understood.
 \vspace{6pt}
 \begin{prop}\label{Prop-defnE-Gateon-state}
 An elementary gate of $k$ qubits $\Lambda^{\theta}_{r}$ sends a basis state of the same number of qubits $\ket{a_1a_2\cdots a_r \cdots a_k}$ to
 \begin{align}\label{eqEgateon-state}
 \Lambda^{\theta}_{r}\ket{a_1a_2\cdots a_r \cdots a_k}
 =\ket{a_1a_2\cdots (a_r+\bm{a}^{\theta}) \cdots a_k},
 \end{align}
 here $r\in[k]$, $\theta=\epsilon_1\epsilon_2\cdots\epsilon_k$ and $\bm{a}^{\theta}=a^{\epsilon_1}_1a^{\epsilon_2}_2\cdots a^{\epsilon_k}_k\in{Z^k_2}$.
 \end{prop}
 \vspace{6pt}
 Apparently, every elementary gate is dimension-one preserving and its own inverse.

 The two propositions reveal the existence of the {\em duality}
 of the {\em polynomial evaluation} and the {\em state computation}.
  \vspace{6pt}
 \begin{cor}\label{Coro-DualityRelation}
 Given a $k$-qubit product operation of elementary gates ${\cal R}$ and its order-reversed product $\hat{{\cal R}}$,
 the equality holds for every basis state $\ket{\bm{x}}$ with $\bm{x}\in{Z^k_2}$,
 \begin{align}\label{eqState-vs-PolyState}
 \ket{{\cal R}\Vdash\bm{x}}=\hat{{\cal R}}\ket{\bm{x}}.
 \end{align}
 \end{cor}
  \vspace{2pt}
 \begin{proof}
 The equality of Eq.~\ref{eqState-vs-PolyState} is deemed as the {\em evaluation duality}  between a state and its associated polynomials.
 Specifically,
 $\ket{{\cal R}\Vdash\bm{x}}=\ket{y_1(\bm{x})y_2(\bm{x})\cdots y_k(\bm{x})}$ stands for a sequence of ordered polynomials written in a state.
 The $s$-th member, $y_s(\bm{x})={\cal R}\Vdash x_s$,
 is the polynomial reaped by acting the product operation ${\cal R}=\Lambda^{\theta_u}_{r_u}\cdots\Lambda^{\theta_2}_{r_2}\Lambda^{\theta_1}_{r_1}$
 embracing $u\geq 1$ elementary gates
 on the $s$-th variable $x_s$ of $\bm{x}=x_1x_2\cdots x_k\in{Z^k_2}$,
 $s\in[k]$.
 The state $\hat{{\cal R}}\ket{\bm{x}}$ is the resultant of activating the order-reversed product
 $\hat{{\cal R}}=\Lambda^{\theta_1}_{r_1}\Lambda^{\theta_2}_{r_2}\cdots\Lambda^{\theta_u}_{r_u}$ of ${\cal R}$ on $\ket{\bm{x}}$.
 This equality elucidates the equivalence of the {\em polynomial evaluation} and the {\em state computation},
 namely $\ket{{\cal R}\Vdash\bm{x}}_{\bm{x}=\bm{a}}=\hat{{\cal R}}\ket{\bm{a}}$
 by substituting a multi-valued string $\bm{a}$ for the input $\bm{x}$ of polynomials $y_s(\bm{x})$ respectively.
 The validness of Eq.~\ref{eqState-vs-PolyState} will be confirmed through the process
 that repetitively employs Eq.~\ref{eqcontrol-onMonomial} to generate polynomial monomials and Eq.~\ref{eqEgateon-state} to calculate state components.

 In the beginning, consider an example of an operation of three gates ${\cal R}=\Lambda^{\theta_3}_{r_3}\Lambda^{\theta_2}_{r_2}\Lambda^{\theta_1}_{r_1}$
 with the order-reversed product $\hat{{\cal R}}=\Lambda^{\theta_1}_{r_1}\Lambda^{\theta_2}_{r_2}\Lambda^{\theta_3}_{r_3}$.
 The two states
 $\ket{\Lambda^{\theta_1}_{r_1}\Vdash\bm{x}}=\ket{\bm{y}^{(1;1)}}$
 and
 $\Lambda^{\theta_3}_{r_3}\ket{\bm{x}}=\ket{\bm{x}^{(3;3)}}$
 have the $s$-components
 \begin{align}\label{eq3gates-genform-one}
 y^{(1;1)}_s=x_s+\delta_{sr_1}\bm{x}^{\theta_1}\hspace{2pt}\text{ and }\hspace{2pt}x^{(3;3)}_s=x_s+\delta_{sr_3}\bm{x}^{\theta_3}
 \end{align}
 educed from acting the first gates of the two operators.
 These two components switch to each other by interchanging the indices $3$ and $1$.
 The applications
 $\ket{\Lambda^{\theta_2}_{r_2}\Lambda^{\theta_1}_{r_1}\Vdash\bm{x}}=\ket{\bm{y}^{(2;1)}}$
 and
 $\Lambda^{\theta_2}_{r_2}\Lambda^{\theta_3}_{r_3}\ket{\bm{x}}=\ket{\bm{x}^{(3;2)}}$ of the second gates
 induce the $s$-th components
 \begin{align}\label{eq3gates-genform-two}
 y^{(2;1)}_s=x_s+\delta_{sr_2}\bm{x}^{\theta_2}+\delta_{sr_1}\prod^k_{t=1}[x_t+\delta_{tr_2}\bm{x}^{\theta_2}]^{\epsilon_{1t}}
 \hspace{2pt}\text{ and }\hspace{2pt}
 x^{(3;2)}_s=x_s+\delta_{sr_3}\bm{x}^{\theta_3}+\delta_{sr_2}\prod^k_{t=1}[x_t+\delta_{tr_3}\bm{x}^{\theta_3}]^{\epsilon_{2t}},
 \end{align}
 $\theta_s=\epsilon_{s1}\epsilon_{s2}\cdots\epsilon_{sk}$ and $s=2,3$.
 The former form of Eq.~\ref{eq3gates-genform-two} converts to the latter by the substitutions of indices $2\rightarrow 3$ and $1\rightarrow 2$,
 and the latter evolves into the former under $3\rightarrow 2$ and $2\rightarrow 1$.
 After exerting the final gates,
 the two states
 $\ket{\Lambda^{\theta_3}_{r_3}\Lambda^{\theta_2}_{r_2}\Lambda^{\theta_1}_{r_1}\Vdash\bm{x}}=\ket{\bm{y}^{(3;1)}}$
 and
 $\Lambda^{\theta_1}_{r_1}\Lambda^{\theta_2}_{r_2}\Lambda^{\theta_3}_{r_3}\ket{\bm{x}}=\ket{\bm{x}^{(3;1)}}$
 procure the identical $s$-th component
  \begin{align}\label{eq3gates-genform-three}
 y^{(3;1)}_s
 &=x_s+\delta_{sr_3}\bm{x}^{\theta_3}+\delta_{sr_2}\prod^k_{t=1}[x_t+\delta_{tr_3}\bm{x}^{\theta_3}]^{\epsilon_{2t}}\notag\\
 &+\delta_{sr_1}\prod^k_{t=1}\{ x_t+\delta_{tr_3}\bm{x}^{\theta_3}+\delta_{tr_2}\prod^k_{t'=1}[x_{t'}+\delta_{t'r_3}\bm{x}^{\theta_3}]^{\epsilon_{t'r_2}} \}^{\epsilon_{1t}}
 =x^{(3;1)}_s.
 \end{align}

 Now, turn to the general case of
 ${\cal R}=\Lambda^{\theta_u}_{r_u}\cdots\Lambda^{\theta_2}_{r_2}\Lambda^{\theta_1}_{r_1}$
 and
 $\hat{{\cal R}}=\Lambda^{\theta_1}_{r_1}\Lambda^{\theta_2}_{r_2}\cdots\Lambda^{\theta_u}_{r_u}$,
 $\theta_i=\epsilon_{i1}\epsilon_{i2}\cdots \epsilon_{ik}\in{Z^k_2}$ and $i\in[u]$.
 In the wake of conducting the first $v\leq u$ gates of ${\cal R}$,
 the $s$-th component of the polynomial state
 $\ket{\Lambda^{\theta_v}_{r_v}\cdots\Lambda^{\theta_2}_{r_2}\Lambda^{\theta_1}_{r_1}\Vdash\bm{x}}=\ket{\bm{y}^{(v;1)}}$
 takes the form
 \begin{align}\label{eqComRonGenForm-x}
 y^{(v;1)}_s=x_s+\delta_{sr_v}\bm{x}^{\theta_v}+\sum^{v-1}_{i=1}\delta_{sr_i}\Gamma_{[v;\hspace{1pt}i]}
 \end{align}
 via Eq.~\ref{eqcontrol-onMonomial}.
 Similarly, subsequent to acting the first $v$ gates of $\hat{{\cal R}}$,
 the $s$-th component of
 $\Lambda^{\theta_{u-v+1}}_{r_{u-v+1}}\cdots\Lambda^{\theta_{u-1}}_{r_{u-1}}\Lambda^{\theta_u}_{r_u}\ket{\bm{x}}=\ket{\bm{x}^{(u;u-v+1)}}$
 is written as
 \begin{align}\label{eqOrderProductRonGenForm-x}
 x^{(u;u-v+1)}_s=x_s+\delta_{sr_u}\bm{x}^{\theta_u}+\sum^{u-1}_{j=u-v+1}\delta_{sr_j}\Gamma_{[u;\hspace{1pt}j]}
 \end{align}
 by Eq.~\ref{eqEgateon-state}.
 Here, the superscript pair $(a;b)$ denotes applying gates in the order from the $b$-th entity up to the $a$-th for $\bm{y}^{(a;b)}$,
 while in the reverse order for $\bm{x}^{(a;b)}$.
 With $e=u$ or $v$ and $1\leq l\leq e-1$,
 every component $\Gamma_{[e;\hspace{1pt}l]}$ reads as
 \begin{align}\label{eqGenFormComponents}
 \Gamma_{[e;\hspace{1pt}l]}=\prod^k_{t_l=1}\{ x_{t_l}+\delta_{t_l,r_e}\bm{x}^{\theta_e} +\sum^{e-l-1}_{q=1}\delta_{t_l,r_q}\Gamma_{[e;e-q]} \}^{\epsilon_{l,t_l}}
 \end{align}
 for $l\neq e-1$ and $\Gamma_{[e;\hspace{1pt}e-1]}=\prod^k_{t=1}\{ x_{t}+\delta_{t,r_e}\bm{x}^{\theta_e}\}^{\epsilon_{e-1,t}}$.
 The form of Eq.~\ref{eqComRonGenForm-x} alters into that of Eq.~\ref{eqOrderProductRonGenForm-x} under the exchanges of indices $v\rightarrow u$ and $i\rightarrow i+(u-v)$.
 Conversely, as per the replacements $u\rightarrow v$ and $j\rightarrow j-(u-v)$,
 Eq.~\ref{eqOrderProductRonGenForm-x} transforms into Eq.~\ref{eqComRonGenForm-x}.
 That is, the changes of indices cause on the conversion of the two equations,
 where the plus-minus sign $\pm$ prefixed to the remainder $u-v$ reflects the opposite gate orders of ${\cal R}$ and $\hat{{\cal R}}$.
 Upon executing the final gates of ${\cal R}$ and $\hat{{\cal R}}$,
 {\em i.e.}, $v=u$,
 there obtain the identities
 $\bm{x}^{\theta_v}=\bm{x}^{\theta_u}$ and $\Gamma_{[v;\hspace{1pt}i]}=\Gamma_{[u;\hspace{1pt}j]}$ for all $i=j$ in Eqs.~\ref{eqComRonGenForm-x} and~\ref{eqOrderProductRonGenForm-x},
 yielding $y^{(u;1)}_s=x^{(u;1)}_s$ for every $s\in[k]$.
 Thus, the corollary is affirmed.
 \end{proof}
 \vspace{6pt}
 It is aesthetic that the succinct formulation of Eq.~\ref{eqState-vs-PolyState}  portrays
 the evaluation equivalence of a polynomial set and the associated state
 under an operation composed of elementary gates.

 Possessing the computational universality,
 elementary gates not only act on quantum states as conventionally known,
 but are also applied on {\em variables} to generate {\em polynomials}.
 This ingenious notion elicits the {\em duality} bridging the polynomial evaluation and the state computation,
 which paves the foundation for encrypting the message and the computation in succeeding sections.

 \section{Message Encryption}\label{secMessageEncrypt}
  \renewcommand{\theequation}{\arabic{section}.\arabic{equation}}
\setcounter{equation}{0}\noindent
 The prospective advancement of quantum-computing technology is anticipated to jeopardize several popular encryptions,
 such as RSA and ECC~\cite{ShorAlg,GroverAlg,2001-Shor-EXP-Chuang,1998-Grover-EXP-Chuang,2023-PQCreview-RB,PostRSA-2017}.
 This has stimulated extensive research on the field of {\em post-quantum} or {\em quantum-resilient cryptography}.
 Of widespread interest in this vein are three types: {\em lattice-based}, {\em code-based} and {\em multivariate-based}
 cryptography~\cite{1988-multivar-MatsuImai,1996-multivar-Patar,2004-multivar-Ding,1998-lattice-HSP,2005-lattice-Regev-LWE,2010-lattice-Regev-RLWE,1978-code-based-McEliece,2008-CodeBased-McESystemAttack,2019-CodeBased-McESystem-review}.
 The security of each cryptosystem is promised by the intractability of solving an {\em NP-complete} problem~\cite{GareyJohnson1979}.
 The lattice-based reckons with searching the shortest lattice vector~\cite{1998-lattice-HSP,2005-lattice-Regev-LWE,2010-lattice-Regev-RLWE},
 the code-based banks on decoding an encrypted message according to its syndrome~\cite{1978-code-based-McEliece,2008-CodeBased-McESystemAttack,2019-CodeBased-McESystem-review}
 and the multivariate-based relies on tackling multivariate nonlinear polynomial equations~\cite{1988-multivar-MatsuImai,1996-multivar-Patar,2004-multivar-Ding}.
 Since the NP-complete problems of the former two types can be respectively rephrased as a task of addressing multivariate
 nonlinear polynomials over an algebraic structure~\cite{2014-MQ=Lattice,2021-MQ=SDP-MPS},
 any new findings in the study of multivariate systems will be of value for the improvements of the duo.
 The multivariate encryption especially holds some potential advantages over the lattice-based.
 First,
 the multivariate-based allows for a faster encoding in terms of simple monomial calculations~\cite{2004-multivar-Ding,2010-lattice-Regev-RLWE}.
 Distinct from lengthy cryptograms over lattices~\cite{2010-lattice-Regev-RLWE},
 compact ciphertexts are acquired,
 each of whose sizes is equal to or slightly larger than that of the plaintext~\cite{2023-MPKC-DeyDutta,2019-Lattice-secure-compare-Bernstein}.
 Third, the decryption is {\em accurate} in contrast to the noisy deciphering of a lattice ciphertext~\cite{2018-decoding-fail-AVV-1089}.
 Note that the code-based keeps the demerits abovesaid
 with relatively less influences~\cite{2016-decode-fail-code-based-GJS-858,2019-CodeBased-McESystem-review}
 but, unlike to the lattice type, has no available scheme of HE.

 The cryptographic primitive of a message will be expounded
 premised on the concepts introduced in the preceding section.
 Briefly, an {\em encryption mapping},
 molded in {\em elementary gates} {\em randomly} chosen,
 is drawn on the generation of a {\em multivariate polynomial set} that is envisaged as the {\em public key} of encryption.
 In virtue of the {\em duality relation},
 a ciphertext is the evaluation of this polynomial set on a plaintext.
 Also acting as the {\em private key},
 the encryption mapping decrypts the ciphertext on account of the duality again and the involutory of elementary gates.
 Ascribable to the {\em invertibility} of elementary gates,
 the decryption is deterministic, {\em i.e.}, {\em exact},
 rather than an erroneous decoding in other cryptosystems~\cite{2009-LatticeCrypt-Regev,2018-decoding-fail-AVV-1089}.
 The protection of a multivariate encryption stems from the hardness of solving polynomial equations~\cite{GareyJohnson1979}.
 EHE is further endowed with a {\em combinatorially} high complexity resistant to quantum adversaries,
 hailing from the difficulty of restoring a {\em circuit} of {\em noncommuting} gates.
 In excess of the standard level $128$ of {\em quantum resilience},
 the security surpasses the suggested threshold $1024$ of {\em hyper quantum resilience}.
 This endeavor is regarded as a concrete realization of {\em noncommutative encryption},
 a concept solely in a theoretical form previously without a tangible construction.

 The algorithm of generating a key pair for the message encryption is articulated as follows.
 The form ${\cal R}\Vdash g(\bm{x})$ stands for the transformation of a multivariate polynomial $g(\bm{x})$
 by applying a product of elementary gates ${\cal R}$
 on variables of each monomial of $g(\bm{x})$ conforming to Eq.~\ref{eqcontrol-onMonomial}.

  \vspace{12pt}
 \noindent
 {\bf Key-Generation Algorithm}

  \noindent
 {\bf Input:} Two positive integers $w$ and $k$, $w\geq k$ \\
 \noindent
 {\bf Output:} The public key,
 $\text{{\rm \textsf{Key}}}_{pub}=\mathscr{P}_{w,\hspace{1pt}k}({\cal R}_{en};\bm{x})$,
 which is an ordered set consisting of a number $w$ of nonlinear polynomials
 of $k$ variables over the binary field $Z_2$,
 and the private key, $\text{{\rm \textsf{Key}}}_{priv}={\cal R}_{en}$,
 which is composed of elementary gates of $k$ qubits
 \begin{enumerate}
 \item
 Prepare an {\em initial} ordered set of polynomials $\mathscr{P}_{in}=\{ g_j(\bm{x}):j\in[w] \}$
 comprising $k$ independent linear polynomials and $w-k$ nonlinear polynomials of low degrees in $k$ variables
 $x_s$ of $\bm{x}=x_1x_2\cdots x_k$, $s\in[k]$.
 \item
 Create an {\em encryption mapping} ${\cal R}_{en}$ that is an {\em ordered product} of {\em nonabelian} elementary gates {\em randomly} chosen.
 \item By applying ${\cal R}_{en}$ on each polynomial of $\mathscr{P}_{in}$,
 output the ordered set of polynomials
 $\text{{\rm \textsf{Key}}}_{pub}=\mathscr{P}_{w,\hspace{1pt}k}({\cal R}_{en};\bm{x})=\{f_j(\bm{x})={\cal R}_{en}\Vdash g_j(\bm{x}):j\in[w]\}$
 serving as the {\em public key}.
 \item Output the {\em private key} ${\cal R}_{en}$.
 \end{enumerate}
 The algorithm favors ${\cal R}_{en}$ including a certain number of multi-controlled gates of higher ranks $\geq 2$
 for the purpose of breeding polynomials of higher degrees in $\mathscr{P}_{w,\hspace{1pt}k}({\cal R}_{en};\bm{x})$.
 Within the composition of ${\cal R}_{en}$,
 a pair of gates $\Lambda^{\theta}_{r}$ and $\Lambda^{\tau}_{s}$ are {\em noncommuting}
 if the $r$-th digit in $\tau$ or the $s$-th digit in $\theta$ is non-null, $r$ and $s\in[k]$.
 To encipher an input message,
 the multivariate polynomial set engendered in step 3 plays the role of a {\em public encryption key},
 or called {\em public key} for short.
 \vspace{6pt}
\begin{prop}\label{defnciphertext}
 Via the public key
 $\mathscr{P}_{w,\hspace{1pt}k}({\cal R}_{en};\bm{x})=\{ f_j(\bm{x}):j\in[w] \}$,
 a $k$-qubit plaintext $\ket{\bm{m}}$ is encoded to the $w$-qubit cipertext
 \begin{align}\label{eqciphertext}
 \ket{\bm{c}}=\ket{f_1(\bm{m})f_2(\bm{m})\cdots f_w(\bm{m})},
 \end{align}
 $\bm{m}\in{Z^k_2}$ and $\bm{c}\in{Z^w_2}$,
 where $f_j(\bm{m})\in{Z_2}$ is the evaluation of the $j$-th polynomial $f_j(\bm{x})\in\mathscr{P}_{w,\hspace{1pt}k}({\cal R}_{en};\bm{x})$.
\end{prop}
\vspace{6pt}
 In simple words, the ciphertext $\ket{\bm{c}}$ is the evaluation of the
 public key $\mathscr{P}_{w,\hspace{1pt}k}({\cal R}_{en};\bm{x})$,
 a multivariate polynomial set, on the input message $\bm{x}=\bm{m}$.

 Founded upon its nature, this message encryption is named the {\em Invertible Multivariate Encryption} and abbreviated as {\em IME}.
 {\em Compact} ciphertexts are permitted hither,
 each of which is sized in the number of polynomials of the public key
 and has the length either a bit larger or even the same as that of the plaintext.
 Whereas, the lattice-based encounters a square ratio of ciphertext-to-plaintext
 that may incur security risks and deteriorated encoding efficiencies~\cite{2019-Lattice-secure-compare-Bernstein}.
 Meanwhile, the key size is effectively reduced in the EHE software
 by compressing each monomial therein and optimizing the memory of storage~\cite{EHE-software-test-web}.
 Leaping over the applicability challenge of concurrency on the lattice encryption with primarily sequential operations,
 the multivariate-based admits inherently vast parallelism attributed to the independent polynomial generation and evaluation,
 {\em cf.} Section~\ref{secExps} for trial records.

 The intricacy of solving a system of polynomial equations~\cite{GareyJohnson1979} provides a safeguard for the multivariate encryption.
 Exclusive of the case $w<k$ of underdefined systems that are easy~\cite{2002-underdefined-MultiVar},
 the algorithm assumes $w\geq k$,
 polynomials outnumbering variables for a public key  $\mathscr{P}_{w,\hspace{1pt}k}({\cal R}_{en};\bm{x})$.
 The occasion that the numbers of polynomials and variables are identical
 is perceived as the hardest~\cite{2002-High-order-XL-alg-NCourt,2017-SolveMQ-fastAlg,2000-SolveMQ-EffAlg-NCourt}.
 In the online EHE~\cite{EHE-software-test-web},
 the number of variables is increased to $w$,
 namely meeting the criterion of the hardest case.
 A dilatation of variable number to raise the security of other encryptions brings on some obstacles
 that compromise the operational efficiencies~\cite{2005-lattice-Regev-LWE,2009-LatticeCrypt-Regev,2004-multivar-Ding}.
 For instance, a quadratic expansion of the ciphertext length occurs in a lattice-based as the security parameter grows~\cite{2019-Lattice-secure-compare-Bernstein}.
 Elevating the number of variables of EHE, however, is relatively straightforward and cost-effective.

 The mechanism of cryptographing a message is analogously employed to encrypt a computation.
 In the next section,
 an ordered set of $n\geq w$ multivariate polynomials is generated from the encrypted action of a desired operation.
 Then, the evaluation of this polynomial set on a ciphertext deduces the sequel of encrypted computation.
 As $n=w$, in particular, both codifications share an identical transformation,
 implying that the message and computation are elegantly mapped to an identical space of encryption.

 The discourse now shifts to the essential strategies of hacking IME.
 First,
 the encryption is defended by the complication of finding a solution for a set of multivariate polynomial equations of degree $\geq 2$,
 which is NP-complete~\cite{GareyJohnson1979,2004-multivar-Ding}.
 The {\em XL method}, a variant of the Groebner-basis approach,
 is recognized as the most effective scheme of attacking polynomial equations~\cite{2017-SolveMQ-Lok,2002-High-order-XL-alg-NCourt,2017-SolveMQ-fastAlg,2000-SolveMQ-EffAlg-NCourt}.
     \vspace{6pt}
 \begin{lem}\label{lemMonomialXL}
 The number of monomials needed in the XL algorithm on a public key
 $\mathscr{P}_{w,\hspace{1pt}k}({\cal R}_{en};\bm{x})$ of degree $d\geq 2$
 is $\sum^{D}_{i=0}{k\choose i}$, here ${k\choose i}$ being a combination number and $d<D\leq k$.
 \end{lem}
 \vspace{2pt}
 \begin{proof}
 The number of monomials involved to solve a set of polynomial equations with the XL algorithm is rendered
 in~\cite{2002-High-order-XL-alg-NCourt,2017-SolveMQ-fastAlg,2000-SolveMQ-EffAlg-NCourt}.
 \end{proof}
 \vspace{6pt}
 If the degree $d$ of $\mathscr{P}_{w,\hspace{1pt}k}({\cal R}_{en};\bm{x})$ as $k\geq 128$ is not small,
 say $d\geq 11$,
 the expended memory to write all necessary monomials is much greater than those of contemporary high-performance computers~\cite{2023-Top500},
 thus failing the algorithm.
 The number of monomials of degrees $\leq D$ is in relation to
 calculating the complexity of solving polynomial equations~\cite{2000-SolveMQ-EffAlg-NCourt,2002-High-order-XL-alg-NCourt}.
   \vspace{6pt}
 \begin{lem}\label{lemXLKeySecurity}
 Breaking a public key
 $\mathscr{P}_{w,\hspace{1pt}k}({\cal R}_{en};\bm{x})$ of degree $d\geq 2$ by the XL algorithm
 requires the complexity ${\cal T}_{XL}=(\sum^{D}_{i=0}{k\choose i})^{\chi}$,
 $d\leq D$ and $2<\chi\leq 3$.
 \end{lem}
 \vspace{2pt}
 \begin{proof}
 Owing to Proposition~\ref{defnciphertext},
 the concerned system consists of $w$ equations $f_j(\bm{m})=c_j$
 from the evaluation of polynomials $f_j(\bm{x})\in\mathscr{P}_{w,\hspace{1pt}k}({\cal R}_{en};\bm{x})$ on $\ket{\bm{c}}=\ket{c_1c_2\cdots c_w}$,
 $j\in[w]$.
 In XL, an extended matrix is formed by treating the monomials of degrees $\leq D$ as independent variables~\cite{2000-SolveMQ-EffAlg-NCourt,2002-High-order-XL-alg-NCourt},
 whose rank equals the number of these monomials $\sum^{D}_{i=0}{k\choose i}$ as in Lemma~\ref{lemMonomialXL}.
 After exercising the Gaussian elimination on this matrix,
 a simplified basis of vectors is produced for solution searching.
 The complexity
 ${\cal T}_{XL}=(\sum^{D}_{i=0}{k\choose i})^{\chi}$ is demanded,
 $2< \chi \leq 3$~\cite{2000-SolveMQ-EffAlg-NCourt,2002-High-order-XL-alg-NCourt}.
 \end{proof}
 \vspace{6pt}
  The constraint $D<<k$ in~\cite{2002-High-order-XL-alg-NCourt}
  is accepted to the estimation of ${\cal T}_{XL}$ for coping with polynomial equations of high degrees.
  Despite the cruciality of degree $D$ in the XL algorithm,
  it is difficult to determine this parameter algorithmically.
  Replacing $D$ by $d$ is suggested here to acquire an underestimated complexity ${\cal T}_{XL}=(\sum^d_{i=0}{k\choose i})^{\chi}$
  for a security measure of the message encryption.
  As $k\geq 128$,
  the complexity ${\cal T}_{XL}$ rises above $2^k$ of the {\em brute-force} method if $k/10\leq d< k/2$.
  Solving quadratic polynomial equations, $d=2$, is still NP-complete~\cite{GareyJohnson1979}.
  By assuming $w\approx k$,
  a subexponential complexity $(\frac{w^{\sqrt{w}}}{\sqrt{w}!})^{\chi}$ is gained for a quadratic system using  XL
  with additional prerequisites, $2<\chi\leq 3$~\cite{2000-SolveMQ-EffAlg-NCourt}.
  This quantity surmounts $2^{128}$ if $w\geq 160$ and $2^{256}$ if $w\geq 400$.
  In a word,
  parameters $w$, $k$ and $d$ are appropriately opted for the criterion of ${\cal T}_{XL}$ exceeding the complexity $2^k$ of brute-force crack.

 Since the private key of IME is a {\em circuit} formed in elementary gates,
 an alternative attack focuses on reconstructing the circuit.
 This attempt is formulated into an inquiry.\\

 \vspace{2pt}
 \noindent{\bf Invertible Circuit Reconstruction Problem, ICRP}\\
 {\em Given the truth table of a multivariate function and two positive integers $t_1$ and $t_2$, decide whether there exists a circuit,
 composed of a maximum of $t_1$ Toffoli gates and $t_2$ negations with at most $t_1$ ancilla bits, in computing this function.}\\

 \noindent
 It will be immediately proved the polynomial equivalence of ICRP to the NP-hard problem below~\cite{2020-MCSP-ILO}.\\

 \vspace{2pt}
 \noindent{\bf Minimum Circuit Size Problem, MCSP}\\
 {\em Given the truth table of a multi-output Boolean function and a positive integer $s$,
 decide whether there exists a Boolean circuit of size no greater than $s$ in computing this function.}\\

 \noindent Here, a Boolean circuit of size $s$ is made up of a number $s$ of classical logic operations of AND, OR and negation.
 The intractability of ICRP is on par with that of MCSP.
 \vspace{6pt}
 \begin{lem}\label{lemMCSP-Reducible}
 ICRP is polynomially equivalent to MCSP.
 \end{lem}
 \vspace{2pt}
 \begin{proof}
 On the basis of the formulation in~\cite{2020-MCSP-ILO},
 consider an arbitrary MCSP instance where the truth table of a function $\beta$ and an upper bound of circuit size $s$ are given.
 Suppose that the circuit in seeking comprises a number $s_1$ of AND, a number $s_2$ of OR and a number $s_3$ of negations,
 $s_1+s_2+s_3\leq s$.
 Resorting to Lemmas~\ref{lemNOTgateQuantum},~\ref{lemANDgateQuantum} and~\ref{lemORgateQuantum},
 the logic operations of the MCSP instance
 are superseded by a number $r_1=s_1+s_2$ of Toffoli gates and a number $r_2=3s_2+s_3$ of negations plus $r_1$ ancilla bits.
 In accordance with this gate substitution,
 the function $\beta$ is translated into the correspondent ICRP instance,
 and it obtains the reduction of MCSP to ICRP.
 The circuit conversion is cost-effective in Toffoli gates and negations,
 rather than a higher expenditure by turning an AND or OR into other elementary gates,
 such as CNOTs and multi-controlled gates of higher ranks.
 Since these Toffoli gates have bits shared in common,
 the number of auxiliary bits is de facto far lower than $r_1$;
 for example,
 by inputting two messages of $l$ bits,
 the circuit of a quantum adder draws on $2$ attached bits
 and that of a quantum multiplier consumes $2l+1$ attendant bits~\cite{QCArithmetic-1st,QCArithmetic-2nd,QCArithmetic-3rd}.

 On the other hand, assume an arbitrary ICRP instance encompassing the truth table of a multivariate function $\gamma$,
 and the upper bounds $t_1$ and $t_2$ of Toffoli gates and negations as well as at most $t_1$ extra bits.
 Exhibited in~\cite{2011-T-gate-replace-SM,2012-XOR-replace-Broesch},
 each toffoli gate is decomposable into three ANDs, one OR and two negations from
 the identity $T^{12}_3={\rm AND}_{12,3}{\rm N}_2{\rm AND}_{23,3}{\rm N}_3{\rm AND}_{23,3}{\rm OR}_{23,3}$
 with qubits indexed $1$, $2$ and $3$.
 Hence, the circuit to be decided for this ICRP instance is rephrased as that of the bounded size $s=6t_1+t_2$ of an MCSP instance,
 which is built exploiting at most $3t_1$ AND, $t_1$ OR and $2t_1+t_2$ negations together with $t_1$ accessory bits.
 The function $\gamma$ is thereby attainable in this MCSP instance.
 It implies that ICRP is reducible to MCSP.
 Therefore, these two problems are polynomially equivalent.
 \end{proof}
 \vspace{6pt}
 Based on this equivalence,
 ICRP receives the complexity measuring of MCSP.
  \vspace{6pt}
 \begin{lem}\label{lemRebuildR-complexity}
 The complexity of solving ICRP in the message encryption of $w$ qubits is a minimum of $2^{w}$.
 \end{lem}
 \vspace{2pt}
 \begin{proof}
 In~\cite{2020-MCSP-ILO}, MCSP has been proven to be harder than the set cover problem,
 whose complexity with $w$ elements is between $w^{O(1)}2^w$ and  $2^{2w}$~\cite{2009-SetCover-BHK,2009-SetCover-Huaetal}.
 Hinged on the polynomial equivalence of the two problems of circuit building,
 the minimum complexity $2^w$ is adopted by neglecting the factor $w^{O(1)}$.
 Albeit ignored here,
 the complexity of preparing a truth table is exponentially high~\cite{2011-discreteMath-Rosen}.
 \end{proof}
 \vspace{6pt}
 With chosen degrees $k/10\leq d<k/2$,
 the complexity ${\cal T}_{XL}$ incurred by the XL algorithm is greater than $2^k$ of the brute-force method.
 Moreover, the complexity ${\cal T}_{ICRP}=2^w$ of solving ICRP overpasses ${\cal T}_{XL}$ if $k\leq w< 2k$,
 {\em i.e.}, ${\cal T}_{ICRP}>{\cal T}_{XL}>2^k$.

 The noncommutativity of elementary gates will offer another strong security.
 The sequent assertion shows that two polynomial sets
 differ from each other if they are generated by products of mutually noncommuting gates in discrepant orders.
 \vspace{6pt}
 \begin{lem}\label{lemmutualNoncomm-MultiCNOT}
 Given a $k$-qubit operator ${\cal Q}$ comprising a number $h\geq 3$ of pairwise noncommuting elementary gates of ranks $\geq 2$
 and a gate-permutation ${\cal Q}_{\sigma}$ of ${\cal Q}$,
 two polynomial sets
 ${\cal P}_{{\cal R}}$
 and
 ${\cal P}_{{\cal R}_{\sigma}}$,
 derived from applying the actions
 ${\cal R}={\cal J}{\cal Q}$ and
 ${\cal R}_{\sigma}={\cal J}{\cal Q}_{\sigma}$ on a set of $k$ independent linear polynomials respectively,
 are not identical,
 where ${\cal J}$ is a product of elementary gates.
 \end{lem}
 \vspace{2pt}
 \begin{proof}
 The proof will show that for a set ${\cal P}=\{ l_r(\bm{x}):r\in[k]\}$ of $k$ independent linear polynomials $l_r(\bm{x})$ of $k$ binary variables,
 the set ${\cal P}_{{\cal R}_{\sigma}}$ precludes a minimum of one polynomial
 ${\cal R}\Vdash l_t(\bm{x})\in{\cal P}_{{\cal R}}$,
 $t\in[k]$.
 Let ${\cal Q}=\Lambda_h\cdots\Lambda_2\Lambda_1=\prod^h_{u=1}\Lambda_u$
 be an ordered product of $h$ mutually noncommuting multi-controlled gates
 of ranks $\geq 2$,
 $\Lambda_u=\Lambda^{\theta_u}_{r_u}$ and $u\in[h]$,
 and $\mathcal{Q}_\sigma=\prod^h_{u=1}\Lambda_{\sigma(u)}$ be the
 permutation of gates in $Q$ under an index transformation $\sigma:[h]\rightarrow[h]$.
 Suppose that the $v$-th gate ${\Lambda}_v$ is the first member moving from the leftmost of ${\cal Q}$ by $\sigma$.
 That is,
 $\Lambda_v$ is mapped to $\Lambda_s$ of $Q_{\sigma}$ for some $s=\sigma(v)$,
 $1\leq s<v\leq h$,
 and ${\cal Q}_{\sigma}$ encloses a number $h-v$ of unchanged gates
 $\Lambda_{\sigma(h)}=\Lambda_h$,
 $\Lambda_{\sigma(h-1)}=\Lambda_{h-1}$,
 $\cdots$,
 $\Lambda_{\sigma(v+1)}=\Lambda_{v+1}$.

 As per Eq.~\ref{eqcontrol-onMonomial},
 it is easy to validate that for every $l_t(\bm{x})\in{\cal P}$,
 $\mathcal{Q}\Vdash l_t(\bm{x})=l_t(\bm{x})+g_t(\bm{x})$
 and
 $\mathcal{Q}_{\sigma}\Vdash l_t(\bm{x})=l_t(\bm{x})+g^{\sigma}_t(\bm{x})$
 have the identical linear polynomial $l_t(\bm{x})$ but different nonlinear entities, $g_t(\bm{x})$ and $g^{\sigma}_t(\bm{x})$.
 Since the gates of ${\cal Q}$ are pairwise noncommuting and $\Lambda_{s+1}$ acts after $\Lambda_s$ in $\mathcal{Q}$,
 $g_t(\bm{x})$ contains a monomial eliminating the variable $x_{s+1}$.
 Provided that $\Lambda_{s+1}$ is placed before $\Lambda_s$ in $\mathcal{Q}_{\sigma}$,
 every monomial of $g^{\sigma}_t(\bm{x})$ must include $x_{s+1}$.
 In other words,
 $g_t(\bm{x})\neq g^{\sigma}_t(\bm{x})$ and ${\cal Q}\Vdash l_t(\bm{x})\neq {\cal Q}_{\sigma}\Vdash l_t(\bm{x})$.
 Together with $l_t(\bm{x})\neq l_e(\bm{x})$,
 the inequality ${\cal Q}\Vdash l_t(\bm{x})\neq {\cal Q}_{\sigma}\Vdash l_e(\bm{x})$ is true as $e\neq t$
 because of discordant linear parts.
 Bottomed on the the early discussion that the inequality holds for $e=t$,
 this result is extended to each integer in $[k]$.
 After multiplying ${\cal J}$ on both sides of the inequality,
 the sequel ${\cal J}{\cal Q}\Vdash l_t(\bm{x})\neq {\cal J}{\cal Q}_{\sigma}\Vdash l_r(\bm{x})$
 is deduced for every $r\in[k]$ and then ${\cal J}{\cal Q}\Vdash l_t(\bm{x})={\cal R}\Vdash l_t(\bm{x})\notin{\cal P}_{{\cal R}_{\sigma}}$.
 Hence, $\mathcal{P}_\mathcal{R}\neq\mathcal{P}_{\mathcal{R}_\sigma}$.
 This also infers that $\mathcal{P}_{{\cal R}_{\sigma}}$ is not an outcome of reordering polynomials in ${\cal P}_{{\cal R}}$.
 \end{proof}
 \vspace{6pt}

 The consequence of Lemma~\ref{lemmutualNoncomm-MultiCNOT} is easily generalized to an operation containing {\em nonabelian} elementary gates.
 \vspace{6pt}
 \begin{thm}\label{thmOPcomplexityMaxset}
 The number of different polynomial sets, generated by all
 permutations of the elementary gates composing an operator ${\cal R}$,
 is a minimum of $h!$,
 where $h$ is the size of a maximal set of pairwise noncommuting gates in ${\cal R}$.
 \end{thm}
 \vspace{2pt}
 \begin{proof}
 It is clear that, for a maximal set assembled by a number $h$ of mutually noncommuting gates of ${\cal R}$,
 an arbitrary product of these $h$ elements must be a
 subproduct of a gate permutation of ${\cal R}$,
 {\em i.e.}, a product containing a less or an equal number of gates of ${\cal R}$.
 Since there are at least $h!$ disparate polynomial sets produced from these products pursuant to Lemma~\ref{lemmutualNoncomm-MultiCNOT},
 the minimum number of polynomial sets generated by all gate permutations of $\mathcal{R}$ is $h!$.
 \end{proof}
 \vspace{6pt}
 As an implication,
 cracking the public key $\mathscr{P}_{w,\hspace{1pt}k}({\cal R};\bm{x})$ generated by an encryption mapping ${\cal R}$,
 whose maximal set of pairwise noncommuting gates is of size $h$,
 costs a {\em combinatorial} complexity comparable to $h!$.
 An encryption mapping ${\cal R}_{en}$ pragmatically often consists of a number $l$ of {\em disjoint} subsets individually created in mutually noncommuting gates
 of size $h_r$, $r\in[l]$.
 Therefore, the {\em decompositional complexity} $h_l!\cdots h_2!h_1!$ is taken to meet the security demand.

 The criterion of complexities to invade the message encryption is propounded.
 \vspace{6pt}
 \begin{cor}\label{CoroComplexityofEncrypt}
 To resolve the encryption with a public key $\mathscr{P}_{w,\hspace{1pt}k}({\cal R}_{en};\bm{x})$ of degree $\geq 2$,
 the following criterion for complexities is imposed that the decompositional noncommutativity complexity ${\cal T}_{deNC}=h_l!\cdots h_2!h_1!$,
 minimally required in retrieving the operator ${\cal R}_{en}$ containing a number $l$ of disjoint products of pairwise noncommuting elementary gates,
 $1\leq \sum^l_{i=1}h_i\leq k$,
 exceeds ${\cal T}_{ICRP}$ of tackling ICRP,
 ${\cal T}_{ICRP}$ surpasses ${\cal T}_{XL}$ of the XL attack,
 and ${\cal T}_{XL}$ is greater than $2^k$ of method de brute force,
 that is, ${\cal T}_{deNC}>{\cal T}_{ICRP}>{\cal T}_{XL}>2^k$.
 \end{cor}
 \vspace{6pt}
 As long as the complexity ${\cal T}_{deNC}$ transcends ${\cal T}_{ICRP}$
 adhering to the conditions $k\geq 128$, $k\leq w< 2k$, $k/10\leq d<k/2$, $l\geq 8$,
 and $k/10\leq h_i< k/2$ for all $i\in[l]$,
 the criterion of Corollary~\ref{CoroComplexityofEncrypt} is always satisfied.
 On this premise,
 IME is safeguarded by the ultimate shield of security ${\cal T}_{deNC}$,
 a {\em combinatorially} high complexity ensuing from an enormous number of possible placements of {\em noncommuting} gates.
 Remark that ${\cal T}_{deNC}$ recommended here is much less than the combinatorial complexity in practical circumstances.

 The encryption conceives the prowess of defying quantum adversaries~\cite{2017-SolveMQ-Lok,2016-Grover-attack-MQ2-West}.
 \vspace{6pt}
 \begin{prop}\label{PropComplexityByGrover}
 The complexity of breaking the public key $\mathscr{P}_{w,\hspace{1pt}k}({\cal R}_{en};\bm{x})$ of degree $\geq 2$
 via the Grover's algorithm is $w^32^{k/2+1}$.
 \end{prop}
 \vspace{6pt}
 The complexity $2^{128}$ minimally installed,
 $k\geq 128$ and $w\geq 160$,
 is decreased to $2^{86}$ if cracking the public key $\mathscr{P}_{w,\hspace{1pt}k}({\cal R}_{en};\bm{x})$ by Grover's algorithm
 on a quantum computer constituted in the present-day fastest quantum gates~\cite{2019-2Qgates-speed-West}.
 Even this reduced complexity spends a temporal course close to the age of universe.
 The security of IME with these parameters is in actuality higher than the level of {\em quantum resilience}.
 This security easily oversteps the threshold $2^{1024}$ as $k\geq 1024$ and $w\geq 1050$,
 which is characterized to be {\em hyper quantum-resilient}.
 An explicit entailment is that the time span of compromising the hyper-quantum resistance approximates the
 eighth power of the duration of breaching the post-quantum standard.

 The key-generation algorithm instructs the decryption by the encryption mapping.
 This mapping emerges as the {\em private encryption key} and abridged as the {\em private key}.
 \vspace{6pt}
 \begin{cor}\label{corDecodingbyKeypriv}
 Encoded from a $k$-qubit plaintext $\ket{\bm{m}}$ in the public key
 $\mathscr{P}_{w,\hspace{1pt}k}({\cal R}_{en};\bm{x})=\{{\cal R}_{en}\Vdash x_j:j\in[w]\}$
 generated by an encryption mapping ${\cal R}_{en}$,
 the $w$-qubit ciphertext $\ket{\bm{c}}$ is decrypted to $\ket{\bm{m}}\otimes\ket{\bm{r}}={\cal R}_{en}\ket{\bm{c}}$ using the private key ${\cal R}_{en}$.
 \end{cor}
 \vspace{2pt}
 \begin{proof}
 Due to the duality,
 the ciphertext $\ket{\bm{c}}=\ket{{\cal R}_{en}\Vdash\bm{x}}_{\bm{x}=\bm{e}}$,
 through evaluating $\mathscr{P}_{w,\hspace{1pt}k}({\cal R}_{en};\bm{x})$ over a $w$-qubit state $\ket{\bm{e}}=\ket{\bm{m}}\otimes\ket{\bm{r}}$,
 equals $\hat{{\cal R}}_{en}\ket{\bm{e}}$.
 Here, $\ket{\bm{r}}$ is a basis state of $w-k$ qubits randomly assigned and
 $\hat{{\cal R}}_{en}$ the order-reversed product of ${\cal R}_{en}$.
 Since every elementary gate is its own inverse,
 $\hat{{\cal R}}^{-1}_{en}={\cal R}_{en}$.
 The plaintext $\ket{\bm{m}}$ is thereby recovered from
 $\hat{{\cal R}}^{-1}_{en}\ket{\bm{c}}={\cal R}_{en}\ket{\bm{c}}=\ket{\bm{e}}$.
 \end{proof}
 \vspace{6pt}
 The proof takes the most difficult case of solving multivariate polynomial equations,
 {\em i.e.}, with an identical number of polynomials and variables.
 In a lattice-based cryptosystem,
 the extra noise during the encryption contributes to the occurrence of decryption failure,
 thus diminishing the security~\cite{2018-decoding-fail-AVV-1089,2016-decode-fail-code-based-GJS-858}.
 By way of contrast, the invertibility of elementary gates brings forth the distinction that the decryption is intrinsically deterministic and thus {\em exact}.
 The computational overhead of this task is proportional to the number of gates composing ${\cal R}_{en}$.
 An encrypted message of appropriate parameters is deciphered within a short time even for an input plaintext of colossal size,
 {\em cf.} Section~\ref{secExps} for the detail.

 Generated by an encryption mapping comprising elementary gates,
 a multivariate polynomial set serves as the public key for the message encryption.
 The ciphertext is derived from the evaluation of this polynomial set on an input plaintext,
 which is decrypted accurately by conducting the encryption mapping.
 Fending off quantum adversaries,
 the security is assured from the intractability of solving multivariate polynomial equations and
 further from a {\em combinatorially} high complexity of restoring the {\em circuit} of the private key.
 The encryption methodology and the security analysis hither are adaptable to cryptifying computations in the succeeding section.

 \section{Cryptovaluation}\label{secHECompute}
  \renewcommand{\theequation}{\arabic{section}.\arabic{equation}}
\setcounter{equation}{0}\noindent
 This section expatiates upon the computation encryption of EHE.
 Remind that a plaintext is encoded into a ciphertext through the polynomial set generated by an {\em encryption mapping} composed of elementary gates.
 Consonantly,
 an {\em encryption transformation} consisting of elementary gates encodes
 a target operation into an encrypted action that is represented as an {\em encrypted-computation polynomial set},
 or simplistically an {\em encrypted polynomial set}.
 Thanks to the {\em duality relation},
 the encrypted computation, termed as {\em cryptovaluation},
 is the fruition of {\em evaluating} the encrypted polynomial set on an input ciphertext.
 In this procedure,
 generating an encrypted polynomial set is time consuming.
 To alleviate workloads,
 the {\em sectional cryptovaluation} is initiated by dividing the {\em circuit} of the encrypted action into a certain number of {\em sections}.
 Since encrypted polynomial sets of the sections are produced {\em independently from one another},
 {\em massive parallelism} is admitted for this generation.

 Dissimilar to approximated homomorphic computations in current HE schemes,
 the cryptovaluation is {\em exact} in virtue of the {\em invertibility} of elementary gates.
 The decryption is {\em accurate} as well
 by applying the {\em private cryptovaluation key} formed in invertible gates,
 rather than the erroneous deciphering of existing cryptosystems.
 Forged with elementary gates enjoying the {\em dimension-one preserving},
 the cryptovaluation is straightforwardly deployable on CPU and GPU environments.
 On account of the obstruction of restoring the {\em circuit}
 from its encrypted polynomial set generated by {\em noncommuting} gates
 and the {\em indiscernibility} of encrypted {\em functions},
 the cryptovaluation is regarded as a {\em blind computation}.
 As a step further,
 the trait of {\em noncommutativity} enables the cryptovaluation not only possessing a security greater than the standard level $128$ for {\em quantum resilience}
 but also crosses the level $1024$ for {\em hyper quantum resilience}.
 Anchored in the successful construction of IME and cryptovaluation,
 the framework EHE is held to be the forefront substantiation of the concept {\em noncommutative cryptography}.

 The establishment of a cryptovaluation is inspired by the concept of Fault Tolerance Quantum Computation (FTQC)
 in Quotient Algebra Partition (QAP)~\cite{SuTsaiQAPFT}.
 Abbreviated as QAPFTQC, this methodology reaches an attaintment that every action in every error-correcting code is fault tolerant~\cite{SuTsaiQAPFT}.
 Literally,
 the quantum code $[n,k,\hspace{2pt}{\cal C}]$ generated by a stabilizer ${\cal C}$ is a QAP~\cite{QAPSu0,QAPSu1,QAPSuTsai1,QAPSuTsai2}.
 Within this structure, a $k$-qubit operation $M$ is encoded into an $n$-qubit {\em fault tolerant encode} of the form
 $U_{en}={\cal A}{\cal T}{\cal M}{\cal B}Q^{\dag}$,
 where ${\cal M}=M\otimes I_{2^{n-k}}$ is the tensor product of $M$ and the $(n-k)$-qubit identity operator
 $I_{2^{n-k}}$, $Q$ an encoding for $[n,k,\hspace{2pt}{\cal C}]$, ${\cal B}$ the correction operator of input errors, ${\cal A}$  the
 correction operator of output errors, and ${\cal T}$ a rotation between input and output errors~\cite{SuTsaiQAPFT}.
 Predicated on this formalism,
 for an arbitrary state $\ket{\psi}$ and its codeword $\ket{\psi_{en}}=Q\ket{\psi}$,
 the $k$-qubit computation $M\ket{\psi}$ is equivalently effectuated by
 its counterpart $U_{en}\ket{\psi_{en}}$ in the encryption space of $n$ qubits,
 referring to~\cite{SuTsaiQAPFT} for the detail.

 The EHE framework borrows the mechanism of QAPFTQC to encipher computations.
 Assume that a $k$-qubit plaintext is encoded into a $w$-qubit ciphertext
 via a multivariate polynomial set generated by an {\em encryption operator} ${\cal R}_{en}$,
 {\em i.e.}, an encryption mapping in last section, $k\leq w$.
 Accompanied by another encryption operator ${\cal R}_{cv}$,
 an $n$-qubit operation $M$, a {\em circuit} of elementary gates,
 is concealed into the encrypted action
 ${\cal U}_{cv}=({\cal R}^{-1}_{en}\otimes I)\hat{M}{\cal R}_{cv}$ with the order-reversed product $\hat{M}$ of $M$,
 $n\geq w$.
 This encrypted action is a simplified form of the fault tolerant encode in QAPFTQC.
 Let the circuit of ${\cal U}_{cv}$ be rephrased as a set of $n$ multivariate polynomials
 following the generating recipe of the preceding two sections.
 Grounded on the poetic duality,
 evaluating this polynomial set on the ciphertext yields the cryptovaluation.
 Finally, ${\cal R}_{cv}$
 serves as the {\em private cryptovaluation key} to decrypt the encrypted computation.

   \begin{figure}[!ht]
 \centering 
 \scalebox{0.5} 
 {\includegraphics{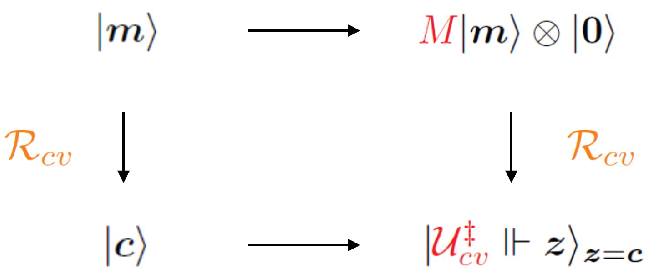}}
 \vspace{12pt} 
 \fcaption{The process of EHE where the message and computation are mapped to an identical space of encryption.~\label{Fig-EncryptCompPoly2Rcv}}
\end{figure}
 \vspace{6pt}
 To begin with, consider $w=n$.
 In this scenario,
 the message and computation are mapped into an identical space of encryption as depicted in Fig.~\ref{Fig-EncryptCompPoly2Rcv}.
  \vspace{6pt}
 \begin{prop}\label{propEncryptCompPoly-RcvRcv}
 Given the $w$-qubit ciphertext $\ket{\bm{c}}$ of a $k$-qubit plaintext $\ket{\bm{m}}$ derived from an encryption operator ${\cal R}_{cv}$ and an $n$-qubit action $M$,
 $n=w\geq k$,
 the cryptovaluation of $M$ on $\ket{\bm{m}}$ is accomplished through the evaluation on $\ket{\bm{c}}$ of the encrypted polynomial set
 \begin{align}\label{eqencryptPoly-subA-Rcv}
 {\cal P}_{n,n}({\cal U}^{\ddag}_{cv};\bm{z})=\{\hspace{2pt} \alpha_i(\bm{z})={\cal U}^{\ddag}_{cv}\Vdash z_i:i\in[n] \hspace{2pt}\}
 \end{align}
 generated by the encrypted action ${\cal U}^{\ddag}_{cv}={\cal R}^{-1}_{cv}\hat{M}{\cal R}_{cv}$,
 where $\hat{M}$ is the order-reversed product of $M$
 and $\alpha_i(\bm{z})$ the $i$-th polynomial of ${\cal P}_{n,n}({\cal U}^{\ddag}_{cv};\bm{z})$, $\bm{z}=z_1z_2\cdots z_n\in{Z^n_2}$.
 \end{prop}
 \vspace{2pt}
 \begin{proof}
 The proof is to disclose the codifications of the message $\ket{\bm{m}}$ and the action $M$ exploiting an encryption operator ${\cal R}_{cv}$.
 Generated by ${\cal R}_{cv}$, the polynomial set $\mathscr{P}_{w,\hspace{1pt}k}({\cal R}_{cv};\bm{x})$,
 the public key for IME,
 encodes $\ket{\bm{m}}$ into a ciphertext $\ket{\bm{c}}$.
 On the strength of the duality relation,
 this ciphertext is alternatively written as $\ket{\bm{c}}=\hat{{\cal R}}_{cv}\ket{\bm{m}}\otimes\ket{\mathbf{0}}$
 from exercising the order-reversed product $\hat{{\cal R}}_{cv}$ of ${\cal R}_{cv}$
 on the product state $\ket{\bm{m}}\otimes\ket{\mathbf{0}}$ of $\ket{\bm{m}}$ and the $(n-k)$-qubit null state $\ket{\bm{0}}$.
 A step further is drawing ${\cal R}_{cv}$ that encodes $M$ into the composition
 $\hat{{\cal U}}^{\ddag}_{cv}=\hat{{\cal R}}_{cv}M\hat{{\cal R}}^{-1}_{cv}$,
 resulting in the encrypted computation $\hat{{\cal U}}^{\ddag}_{cv}\ket{\bm{c}}=\hat{{\cal R}}_{cv}M\ket{\bm{m}}\otimes\ket{\mathbf{0}}$ called the {\em cryptovaluation}.
 Here, $\hat{{\cal U}}^{\ddag}_{cv}$ is the order-reversed product of the encrypted action ${\cal U}^{\ddag}_{cv}$.

 With the associated state
 $\ket{{\cal U}^{\ddag}_{cv}\Vdash\bm{z}}=\ket{\alpha_1(\bm{z})\alpha_2(\bm{z})\cdots\alpha_n(\bm{z})}$,
 $\alpha_i(\bm{z})={\cal U}^{\ddag}_{cv}\Vdash z_i$
 and
 $i\in[n]$,
 it relishes the duality
 $\hat{{\cal U}}^{\ddag}_{cv}\ket{\bm{c}}=\ket{{\cal U}^{\ddag}_{cv}\Vdash\bm{z}}_{\bm{z}=\bm{c}}$
 between the state computation and the polynomial evaluation.
 Thus, the cryptovaluation is engaged in $\ket{{\cal U}^{\ddag}_{cv}\Vdash\bm{z}}_{\bm{z}=\bm{c}}$ of calculating the polynomial set
 ${\cal P}_{n,n,p}({\cal U}^{\ddag}_{cv};\bm{z})$ on the ciphertext $\ket{\bm{c}}$.
 The operator $\hat{{\cal R}}^{-1}_{cv}={\cal R}_{cv}$
 works as the {\em private cryptovaluation key} of the decryption,
 namely
 ${\cal R}_{cv}\ket{{\cal U}^{\ddag}_{cv}\Vdash\bm{z}}_{\bm{z}=\bm{c}}
 ={\cal R}_{cv}\hat{{\cal U}}^{\ddag}_{cv}\ket{\bm{c}}=M\ket{\bm{m}}\otimes\ket{\mathbf{0}}$.
 Refer to Fig.~\ref{Fig-EncryptCompPoly2Rcv} for the diagram outlining the process.
 In the scenario $n=w$,
 the message and computation are elegantly sent into an identical space of encryption under the same encryption operator ${\cal R}_{cv}$.
 \end{proof}
 \vspace{6pt}

\begin{figure}[!ht]
 \centering 
 \scalebox{0.5} 
 {\includegraphics{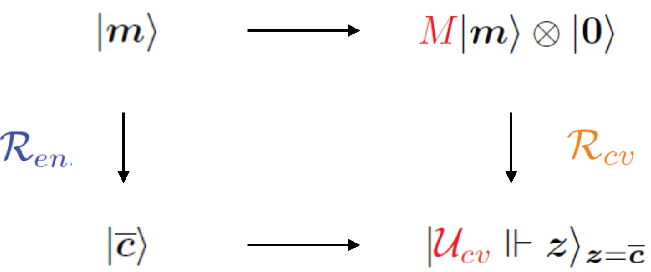}}
 \vspace{12pt} 
 \fcaption{The process of EHE that the message and computation are mapped to different spaces of encryption.~\label{Fig-EncryptCompPolyRcvRen}}
\end{figure}

 The formulation of Proposition~\ref{propEncryptCompPoly-RcvRcv} is less efficient because,
 in comparison with IME using $w$ qubits,
 it adds extra $n-w$ ancilla qubits.
 For better performances,
 the message and computation are encoded resorting to two different encryption operators as displayed in Fig.~\ref{Fig-EncryptCompPolyRcvRen}.
 \vspace{6pt}
 \begin{prop}\label{propEncryptCompPoly-RcvRen}
 Given the $w$-qubit ciphertext $\ket{\bm{c}}$ of a $k$-qubit plaintext $\ket{\bm{m}}$
 derived from an encryption operator ${\cal R}_{en}$ and an $n$-qubit action $M$, $n>w\geq k$,
 the cryptovaluation of $M$ on $\ket{\bm{m}}$
 is accomplished through the evaluation on $\ket{\bm{c}}$ of the encrypted polynomial set
 \begin{align}\label{eqencryptPoly-subA-Rcv}
 {\cal P}_{n,w}({\cal U}_{cv};\bm{z})=\{\hspace{2pt} \beta_i(\bm{z})={\cal U}_{cv}\Vdash z_i:i\in[n] \hspace{2pt}\}
 \end{align}
 generated by the encrypted action ${\cal U}_{cv}=({\cal R}^{-1}_{en}\otimes I)\hat{M}{\cal R}_{cv}$,
 where ${\cal R}_{cv}$ is an $n$-qubit encryption operator,
 $\beta_i(\bm{z})$ the $i$-th polynomial of ${\cal P}_{n,w}({\cal U}_{cv};\bm{z})$ with $\bm{z}=z_1z_2\cdots z_n\in{Z^n_2}$
 and $I$ the identity operator of $n-w$ qubits.
 \end{prop}
 \vspace{2pt}
 \begin{proof}
 This proposition is analogous to the concept of QAPFTQC creating fault tolerant encodes of $M$~\cite{SuTsaiQAPFT}.
 Here, the encode
 $\hat{{\cal U}}_{cv}=\hat{{\cal R}}_{cv}M(\hat{{\cal R}}^{-1}_{en}\otimes I)$
 is the order-reversed product of encrypted action ${\cal U}_{cv}$,
 with $M$ sandwiched by the operator of input errors $\hat{{\cal R}}^{-1}_{en}\otimes I$ and the operator of output errors $\hat{{\cal R}}_{cv}$.
 The proof follows the same procedure as in Proposition~\ref{propEncryptCompPoly-RcvRcv}
 but replacing the encryption operator ${\cal R}^{-1}_{cv}$ of ${\cal U}^{\ddag}_{cv}$ by ${\cal R}^{-1}_{en}\otimes I$,
 the encrypted polynomial set ${\cal P}_{n,n}({\cal U}^{\ddag}_{cv};\bm{z})$
 by ${\cal P}_{n,w}({\cal U}_{cv};\bm{z})$,
 and the polynomial state $\ket{{\cal U}^{\ddag}_{cv}\Vdash\bm{z}}$ by $\ket{{\cal U}_{cv}\Vdash\bm{z}}$.
 Similarly,
 ascertained from the duality relation,
 the output of the cryptovaluation is the polynomial evaluation $\ket{{\cal U}_{cv}\Vdash\bm{z}}_{\bm{z}=\overline{\bm{c}}}$
 on the product state $\ket{\overline{\bm{c}}}=\ket{\bm{c}}\otimes\ket{\mathbf{0}'}$ of $\ket{\bm{c}}$ and a null basis state $\ket{\mathbf{0}'}$ of $n-w$ qubits.
 Likewise, the operator ${\cal R}_{cv}$ decrypts the evaluation.
 The diagram of Fig.~\ref{Fig-EncryptCompPolyRcvRen} pictures this process.
 \end{proof}
 \vspace{6pt}
 In accordance with Propositions~\ref{propEncryptCompPoly-RcvRcv} and~\ref{propEncryptCompPoly-RcvRen},
 the number of qubits $n$ is prescribed to accommodate sufficiently numerous and sophisticated computations,
 which embrace a necessary number of ancilla bits in order to cover all desired actions.
 The requisite $n>2k$ offers the freedom of {\em indiscernibility} amongst
 the diverse functions as in Section~\ref{secExps}.
 While, this requisite is always modifiable to support more types of functions that are unidentified during the entire course.
 It is worth noting that based on the independent evaluation of polynomials and the inherent configurability of circuit implementation for encoded functions,
 the framework EHE is seamlessly applicable to the
 {\em secure multi-party computation}~\cite{1986-multi-party-compute-Yao,1987-multi-party-compute-GMW,2018-multi-party-compute-EKR-etal,2019-multi-party-compute-Zhao-etal,2008-CompNetwork-Kung}.
  \vspace{6pt}
 \vspace{0pt}

 Generating an encrypted polynomial set is a particularly time-intensive task throughout the cryptovaluation,
 yet, to be praised, it welcomes treatments in parallel.
 Albeit the {\em sequential} evaluation of the cryptovaluation,
 the {\em circuit} of an encrypted action admits a division of multiple {\em sections}.
 On the basis of the {\em independent} production of encrypted polynomial sets from the sections,
 there wins {\em vast parallelism} in the polynomial generation.
 This approach, called {\em sectional cryptovaluation},
 is thus advantageous for inducing efficient encrypted computations.
 The following assertion orchestrates this strategy in respect of Proposition~\ref{propEncryptCompPoly-RcvRen},
 which is valid to Proposition~\ref{propEncryptCompPoly-RcvRcv} also.
 \vspace{6pt}
 \begin{prop}\label{propSectionEncryption}
 Given an encrypted action ${\cal U}_{cv}$ of an operation $M$,
 the sectional cryptovaluation of $M$ is accomplished through the sequential evaluation of encrypted polynomial sets
 \begin{align}\label{eqencryptSeqPolySetEvaluate}
 {\cal P}_{n,w}({\cal U}_{cv,\hspace{1pt}q};\bm{z})=\{\hspace{2pt} \beta_{i,\hspace{1pt}q}(\bm{z})={\cal U}_{cv,\hspace{1pt}q}\Vdash z_i:i\in[n] \hspace{2pt}\}
 \end{align}
 that are generated in massively parallel from a number $e$ of sectional encrypted circuits ${\cal U}_{cv,\hspace{1pt}q}$ composing ${\cal U}_{cv}$, $q\in[e]$,
 where $\beta_{i,\hspace{1pt}q}(\bm{z})$ is the $i$-th polynomial of ${\cal P}_{n,w}({\cal U}_{cv,\hspace{1pt}q};\bm{z})$.
 \end{prop}
 \vspace{2pt}
 \begin{proof}
 The key is to firstly divide the circuit of encrypted action ${\cal U}_{cv}$ into a number $e$ of sections.
 Depending on computing environments,
 the number $e$ ranges from $n/2$ to $4n$ on the single-CPU and from $n/8$ to $n$ on the multiple cores~\cite{test-environments}.
 Due to this division, the circuit is factorized into a product ${\cal U}_{cv}={\cal U}_{e}{\cal U}_{e-1}\cdots{\cal U}_2{\cal U}_1$ of $e$ component actions ${\cal U}_q$,
 $q\in[e]$.
 By arbitrarily taking a number $e$ of {\em sectional encryption operators} ${\cal R}_q$ individually comprising elementary gates randomly generated,
 each member ${\cal U}_q$ is converted into a {\em sectional encrypted circuit} ${\cal U}_{cv,\hspace{1pt}q}={\cal R}_q{\cal U}_q{\cal R}^{-1}_{q-1}$
 for $2\leq q\leq e-1$, with ${\cal U}_{cv,1}={\cal R}_1{\cal U}_1$ and ${\cal U}_{cv,\hspace{1pt}e}={\cal U}_{e}{\cal R}^{-1}_{e}$.
 That is, the encrypted action is rewritten as
 ${\cal U}_{cv}
 =({\cal U}_e{\cal R}^{-1}_e)({\cal R}_e{\cal U}_{e-1}{\cal R}^{-1}_{e-1})\cdots({\cal R}_2{\cal U}_2{\cal R}^{-1}_1)({\cal R}_1{\cal U}_1)
 ={\cal U}_{cv,e}{\cal U}_{cv,e-1}\cdots {\cal U}_{cv,2}{\cal U}_{cv,1}$.

 Since encrypted polynomial sets ${\cal P}_{n,w}({\cal U}_{cv,\hspace{1pt}q};\bm{z})$ are engendered independently from the encrypted circuits ${\cal U}_{sv,q}$,
 it enables a highly concurrent generation of polynomial sets.
 Founded on the duality,
 the sequential evaluation of polynomial states $\ket{{\cal U}_{cv,\hspace{1pt}q}\Vdash\bm{z}}$ educes the harvest of the cryptovaluation.
 With the initial input $\ket{\overline{\bm{c}}}$ in the proof of Proposition~\ref{propEncryptCompPoly-RcvRen},
 a prior output is tapped as the subsequent input of steps from $q'=1$ to $q'=e$,
 {\em i.e.},
 $\ket{\overline{\bm{c}}_1}=\ket{{\cal U}_{cv,\hspace{1pt}1}\Vdash\bm{z}}_{\bm{z}=\overline{\bm{c}}}$
 and
 $\ket{\overline{\bm{c}}_{q'+1}}=\ket{{\cal U}_{cv,\hspace{1pt}q'+1}\Vdash\bm{z}}_{\bm{z}=\overline{\bm{c}}_{q'}}$
 for $q'< e$.
 The final evaluation $\ket{\overline{\bm{c}}_e}$ is the consequent of this encrypted computation.
 \end{proof}
 \vspace{6pt}
 As entities accessible openly,
 the encrypted polynomial sets are tantamount to the {\em public cryptovaluation key} for cryptographing a computation.
 The development of EHE software capitalizes on the sectional cryptovaluation,
 referring to trial records in Section~\ref{secExps}.

 Hinged on previous propositions,
 an operation is encoded into an encrypted action that is represented as a single or several publicly-accessible encrypted polynomial sets.
 The security of a cryptovaluation thus arises out of the obstacle of constructing the {\em circuit} of the encrypted action.
   \vspace{6pt}
 \begin{cor}\label{corSecurityEncryptCompPoly-Ren}
 The complexity of invading an $n$-qubit cryptovaluation on a $w$-qubit ciphertext is greater than $2^{w}$, $w\leq n$.
 \end{cor}
 \vspace{2pt}
 \begin{proof}
 In Propositions~\ref{propEncryptCompPoly-RcvRcv} and~\ref{propEncryptCompPoly-RcvRen},
 the $n$-qubit cryptovaluation is attacked by restoring the {\em circuit} of encrypted action of an operation,
 which returns to ICRP.
 The complexity of solving ICRP for an $n$-qubit system outstrips $2^n$ according to Lemma~\ref{lemRebuildR-complexity}.
 As $n\geq w$, the minimum complexity $2^w$ is retained.
 The cryptovaluation fails if invaders successfully rebuild the encryption operators.
 Since these two operators are identical in Proposition~\ref{propEncryptCompPoly-RcvRcv} and are different in Proposition~\ref{propEncryptCompPoly-RcvRen},
 the complexity of retrieving them of the latter is greater than that of the former.
 Yet, this complexity is waived here.
 Breaking the sectional cryptovaluation of Proposition~\ref{propSectionEncryption} is to recover the encrypted circuit ${\cal U}_{cv,\hspace{1pt}q}$ of each section,
 {\em i.e.}, solving ICRP again,
 which necessitates a complexity surmounting $2^n$.
 Likewise, it procures the minimum $2^w$ only.
 Even though the encryption operators of ${\cal U}_{cv,\hspace{1pt}q}$ are distinct,
 they are restored similarly to those of Proposition~\ref{propEncryptCompPoly-RcvRen},
 but as well with no regard for the corresponding complexity.
 \end{proof}
 \vspace{6pt}
 The security of a cryptovaluation should fulfill the criterion of complexities as in Corollary~\ref{CoroComplexityofEncrypt},
 that is, ${\cal T}_{deNC}>{\cal T}_{ICRP}>{\cal T}_{XL}>2^k$.
 To meet this criterion,
 the choices of parameters $k\geq 128$, $k\leq w<2k$, $k/10\leq d< k/2$, $l\geq 8$ and
 $k/10\leq h_i\leq k/2$ for all $i\in[l]$ are reckoned with.
 Here, $l$ is the number of disjoint products of mutually {\em noncommuting} elementary gates of an encrypted action,
 and $h_i$ is the number of  gates of the $i$-th product.
 Noteworthily, ${\cal T}_{deNC}$ is a {\em combinatorially} high complexity
 of deciding proper choices from a gigantic number of possible arrangements of noncommuting gates,
 which is much underestimated compared with the target measure of security.
 Forasmuch as the difficulty of reinstating an operation from its encrypted polynomial set and the indiscernibility of enciphered functions,
 the cryptovaluation is viewed as a {\em blind} computation,
 a feature absent from HE in existence.
 Under the crack of Grover's algorithm~\cite{GroverAlg,2016-Grover-attack-MQ2-West},
 the complexity ${\cal T}_{ICRP}=2^w$ is reduced to $2^{w/2}$,
 still devoting a time close to the age of universe.
 Respecting the parameters abovestated,
 the security is greater than the standard of {\em quantum resilience} as $k\geq 128$ and $w\geq 160$,
 and forwardly transcends the suggested threshold $2^{1024}$ of {\em hyper-quantum resilience}
 for $k\geq 1024$ and $w\geq 1050$.

 Pursuant to the {\em invertibility} of elementary gates,
 the cryptovaluation of EHE is {\em exact},
 surpassing approximate homomorphic computations elicited by reducing the noise that quadratically accumulates per operation~\cite{2018-AAAC-HE-Review,HE-BGV-2011,HE-GHS-2012,2019-HE-standard}.
 In order to control a such noise growth of a two-fold expanded ciphertext~\cite{2018-HE-Review-RL,2019-HE-standard},
 current HE schemes constrain the types of concerned operations and computation sizes~\cite{HE-RLWE-2017-Kerskin,HE-BGV-2011,HE-GHS-2012,2019-HE-standard}.
 However, the cryptovaluaion allows for a compact ciphertext encoded from a plaintext of extensive size,
 with no limitation on the types of encrypted functions.
 The cryptovalution is further effectively enhanced through massive parallelism of begetting and evaluating independent monomials,
 rather than the known HE that encounters sequential cryptified computations ill-suited to concurrency~\cite{HE-BGV-2011,HE-GHS-2012,2019-HE-standard}.

  Experimental calculation outcomes of the encrypted {\em elementary functions} will be presented in Section~\ref{secExps}.
  Pragmatically,
  three essential costs should be addressed.
  The workload of generating encrypted polynomial sets from the divided sections is rendered at first.
   \vspace{6pt}
 \begin{cor}\label{corExpenseofPrepareCryptovaluation-Poly}
 Towards establishing a sectional cryptovaluation of the elementary function of $n$ qubits,
 the expense of generating encrypted polynomial sets is proportional to $\kappa n$,
 $n^2\leq \kappa\leq n^3$.
 \end{cor}
 \vspace{2pt}
 \begin{proof}
 For the sake of indiscernibility of different operations to fulfill the blind computation,
 the runtimes of generating encrypted polynomial sets of these functions are arranged to be close.
 Here, the number $e$ of sections is opted in the range $n/2\leq e\leq 4n$ on the single-CPU and $n/8\leq e\leq n$ on multiple cores~\cite{test-environments}.
 By means of immense parallelism of the generation,
 the expense counts the complexity of preparing a polynomial set of a single section,
 which amounts to the product $\kappa n$ of the number $\kappa$ of steps for deducing a polynomial and the number $n$ of polynomials.
 As the concerned polynomial, the longest member of the set,
 containing at most $n^2$ monomials in practice,
 the number of steps $\kappa$ for creating monomials and eliminating duplicate members of the polynomial
 is within the maximum, $n\cdot n^2=n^3$, thus $n^2\leq \kappa\leq n^3$.
 \end{proof}
 \vspace{6pt}
 If another set of functions is designated,
 in order to retain the blind computation,
 the number of sections may be adjusted to effectuate similar exercised times of generating polynomial sets.

 The total overhead of evaluating polynomial sets for a sectional cryptovalaution is the accumulated consumption from distinct sections.
   \vspace{6pt}
 \begin{cor}\label{corCostofcryptovaluation-Poly}
 The computational cost of a sectional cryptovaluation, summing sectional costs,
 is  proportionally bounded by the monomial number of the longest polynomial in all sections.
 \end{cor}
 \vspace{2pt}
 \begin{proof}
 This assertion is a consequence of the sequential evaluation of the sectional cryptovaluation.
 \end{proof}
 \vspace{6pt}
 Remind that the number $e$ of sections falls in the interval $n/2\leq e\leq 4n$ on the single-CPU and $n/8\leq e\leq n$ on multiple cores.
 For the maximum monomial number $n^2$ as per test data,
 the overall cost proportionally ranges from $n^3/8$ to $4n^3$.

 The deciphering of a cryptovalaution is to apply the {\em private cryptovaluation key},
 a product of elementary gates, on the final output state from a sequential evaluation.
  \vspace{6pt}
 \begin{cor}\label{corDecryptCryptovaluation-Poly}
 The computational cost of decrypting a cryptovaluation is proportional to the number of elementary gates instituting its private cryptovaluation key.
 \end{cor}
 \vspace{2pt}
 \begin{proof}
 This is true because the private cryptovaluation key is a {\em circuit} consisting of elementary gates,
 referring to Propositions~\ref{propEncryptCompPoly-RcvRcv},~\ref{propEncryptCompPoly-RcvRen} and~\ref{propSectionEncryption}.
 \end{proof}
 \vspace{6pt}
 The gate number of a private cryptovaluation key is realistically linear in $n$ for an $n$-qubit cryptovaluation.
 The decryption of an existent HE may bump into a failure probability caused by noise,
 which grows with the problem size and lowers the security level as well as the efficiency~\cite{2019-HE-standard,2018-decoding-fail-AVV-1089}.
 Whereas, predicated on a direct application of elementary gates on the encrypted state,
 the decryption is {\em exact} and cost-effective.

 \section{Experimental Finding}\label{secExps}
  \renewcommand{\theequation}{\arabic{section}.\arabic{equation}}
\setcounter{equation}{0}\noindent
 The EHE software consists of two codes in the $64$-bit computing architecture:
 one for IME only and the other for executing EHE inclusive of both encryptions of the message and computation.
 These two codes are deployed on three environments of
 the single-CPU, multi-CPUs and single-node
 GPU~\cite{test-environments}.
 Among the {\em elementary functions} for experiments,
 the addition, subtraction and string comparison respectively comprise a number of elementary gates linear in $k$~\cite{QCAddition-Ripple,QCAddition-Taka,QCArithmetic-1st},
 and the multiplication, division, sum of squares and monomial powers are individually constituted with gates numbered in $k^2$~\cite{QCArithmetic-2nd,QCArithmetic-3rd,QCMulti-Karatsuba}.
 The parameters $k$, $w$, $n$ and $d$ here are chosen appropriately to satisfy the security criteria in
 Corollaries~\ref{CoroComplexityofEncrypt} and~\ref{corSecurityEncryptCompPoly-Ren}.

   \vspace{0pt}
  \begin{table}[!ht]
 \small
\[ \begin{array}{cccccccc}
                          &  t_{kg\text{-}sc} &  t_{en\text{-}sc}
                          &  t_{kg\text{-}mc} &  t_{en\text{-}mc}
                          &  t_{kg\text{-}sg} &  t_{en\text{-}sg}
                          &
                           \begin{aligned}
                             &\hspace{20pt}t_{de\text{-}sc} \\
                             &(t_{de\text{-}mc},t_{de\text{-}sg})
                           \end{aligned}
\\
                          &                   &                   &
                          &                   &                   &
\\
 \vspace{2pt}(\hspace{4.5pt}128\hspace{0pt},\hspace{5pt}160)
                          &   2.5\sim 22 \hspace{1pt}\text{s}       & 0.2\sim 1.8 \hspace{1pt}\text{s}
                          & 0.15\sim 2 \hspace{1pt}\text{s}        & 10^{-3} \hspace{1pt}\text{s}
                          & 0.5 \hspace{1pt}\text{s}              &10^{-3} \hspace{1pt}\text{s}
                          & 0.002 \hspace{1pt}\text{s}
 \\
 \vspace{2pt}(\hspace{4.5pt}256\hspace{0pt},\hspace{5pt}280)
                          &   12\sim 45   \hspace{1pt}\text{s}    &1.4\sim 3.6 \hspace{1pt}\text{s}
                          & 1.5\sim 3.5 \hspace{1pt}\text{s}      & 10^{-3} \hspace{1pt}\text{s}
                          & 2.5 \hspace{1pt}\text{s}              & 10^{-2} \hspace{1pt}\text{s}
                          & 0.002 \hspace{1pt}\text{s}
 \\
 \vspace{2pt}(\hspace{4.5pt}512\hspace{0pt},\hspace{5pt}540)
                          &   48\sim 145  \hspace{1pt}\text{s}    & 2.8\sim 4.8 \hspace{1pt}\text{s}
                          &   5\sim 12  \hspace{1pt}\text{s}    & 10^{-2} \hspace{1pt}\text{s}
                          &   10  \hspace{1pt}\text{s}            & 10^{-2} \hspace{1pt}\text{s}
                          &   0.05 \hspace{1pt}\text{s}\\
 \vspace{2pt}(1024,1050)  &   170\sim 300 \hspace{1pt}\text{s}    & 7\sim 16 \hspace{1pt}\text{s}
                          &   4\sim 9 \hspace{1pt}\text{s}       & 0.9 \hspace{1pt}\text{s}
                          &   15 \hspace{1pt}\text{s}             &   1.5 \hspace{1pt}\text{s}
                          &   0.08 \hspace{1pt}\text{s}\\
 \vspace{2pt}(1536,1560)  &   295\sim 480 \hspace{1pt}\text{s}    & 17\sim 28\hspace{1pt}\text{s}
                          &   13\sim 42  \hspace{1pt}\text{s}     & 2\hspace{1pt}\text{s}
                          &   45  \hspace{1pt}\text{s}            & 2.2 \hspace{1pt}\text{s}
                          &0.1\hspace{1pt}\text{s}\\
 \vspace{2pt}(2048,2080)  &  \text{N/A}       & \text{N/A}
                          &  270\sim 340 \hspace{1pt}\text{s}     & 20 \hspace{1pt}\text{s}
                          &  450 \hspace{1pt}\text{s}             & 25 \hspace{1pt}\text{s}
                          &  0.4 \hspace{1pt}\text{s} \\
 \vspace{2pt}(4096,4160)  &  \text{N/A}       & \text{N/A}
                          &  25\sim 42 \hspace{1pt}\text{mins}    & 80 \hspace{1pt}\text{s}
                          &  \text{N/A}       &\text{N/A}
                          &  1.2\hspace{1pt}\text{s}\\
 \vspace{2pt}(6400,6440)  &  \text{N/A}       & \text{N/A}       &  45\sim 60 \hspace{1pt}\text{mins}       & 180\hspace{1pt}\text{s}
                          &  \text{N/A}       &\text{N/A}
                          &  1.8\hspace{1pt}\text{s}\\
\end{array}\]
 \vspace{12pt} 
 \tcaption{Performance of IME on the three environments.~\label{Fig-bks-MessageE-laptop-HPC}}
\end{table}
 \vspace{0pt}

 To conveniently show the test data of the message encryption,
 the two parameters of the public key $\mathscr{P}_{w,\hspace{1pt}k}({\cal R}_{en};\bm{x})$ are put into the pair $(k,w)$.
 As in Table~\ref{Fig-bks-MessageE-laptop-HPC},
 $t_{kg\text{-}sc}$, $t_{kg\text{-}mc}$ and $t_{kg\text{-}sg}$ denote the key-generation times
 respectively on the single-CPU, multi-CPUs and single-node GPU~\cite{test-environments},
 $t_{en\text{-}sc}$, $t_{en\text{-}mc}$ and $t_{en\text{-}sg}$ the encoding times,
 and $t_{de\text{-}sc}$, $t_{de\text{-}mc}$ and $t_{de\text{-}sg}$ the deconding times.
 The duration of reading and exchange of data is absorbed,
 which occupies around $4\%$ in the key generation, $90\%$ in the encoding and $2\%$ in the deconding.
 Imputed to the relatively vast parallelism, greater amount of memory and faster data transfer,
 the platforms of multi-CPU and single-node GPU earn an approximately tenfold to twentyfold increase for the efficiencies of key generation and encoding,
 compared with the single-CPU.
 The time frames of decoding, composed of elementary gates numbered linearly in $w$,
 operating on ciphertexts are short and close on the three systems.
 The case up to the maximum of parameter pair $(6400,6440)$ is brought forward,
 which is associated with an ample encryption of high security not easy to arrive at for post-quantum cryptosystems in existence.
 Implemented within reasonable time increments of the key generation and the encoding,
 the sectional stratagem is equally well-adapted for the message encryption,
 called {\em multi-layer} IME~\cite{EHE-software-test-web},
 and is anticipated to further heighten the level of security.

    \vspace{0pt}
  \begin{table}[!ht]
 \small
\[ \begin{array}{cccccccc}
                          &  T_{kg\text{-}sc} &  T_{evl\text{-}sc}
                          &  T_{kg\text{-}mc} &  T_{en\text{-}mc}
                          &  T_{kg\text{-}sg} &  T_{en\text{-}sg}
                          &
                           \begin{aligned}
                             &\hspace{20pt}T_{de\text{-}sc} \\
                             &(T_{de\text{-}mc},T_{de\text{-}sg})
                           \end{aligned}
\\
                          &                   &                   &
                          &                   &                   &
\\
 \vspace{2pt}(\hspace{4.5pt}128\hspace{0pt},\hspace{5pt}160\hspace{0pt},\hspace{5pt}240)
                          &   10 \hspace{1pt}\text{mins}       & 30 \hspace{1pt}\text{s}
                          &   1 \hspace{1pt}\text{min}         & 2 \hspace{1pt}\text{s}
                          &   1 \hspace{1pt}\text{min}          &   3 \hspace{1pt}\text{s}
                          & 0.002 \hspace{1pt}\text{s}
 \\
 \vspace{2pt}(\hspace{4.5pt}256\hspace{0pt},\hspace{5pt}280\hspace{0pt},\hspace{5pt}400)
                          &  21 \hspace{1pt}\text{mins}       & 4 \hspace{1pt}\text{mins}
                          & 2 \hspace{1pt}\text{mins}         &  12 \hspace{1pt}\text{s}
                          &  2.3 \hspace{1pt}\text{mins}                &  13 \hspace{1pt}\text{s}
                          &  0.03 \hspace{1pt}\text{s}
 \\
 \vspace{2pt}(\hspace{4.5pt}512\hspace{0pt},\hspace{5pt}540\hspace{0pt},\hspace{5pt}750)
                          &  \text{N/A}       & \text{N/A}
                          &  8 \hspace{1pt}\text{mins}         &  1.8 \hspace{1pt}\text{mins}
                          &  9 \hspace{1pt}\text{mins}         &  2 \hspace{1pt}\text{mins}
                          &  0.06 \hspace{1pt}\text{s}\\
 \vspace{2pt}(1024,1050,1600)  &  \text{N/A}       & \text{N/A}
                               &   20 \hspace{1pt}\text{mins}           & 4 \hspace{1pt}\text{mins}
                               &  21 \hspace{1pt}\text{mins}            & 4.2 \hspace{1pt}\text{mins}
                               &0.1 \hspace{1pt}\text{s}\\
 \vspace{2pt}(1536,1560,2400)  &  \text{N/A}       & \text{N/A}
                               &   32 \hspace{1pt}\text{mins}           & 6 \hspace{1pt}\text{mins}
                               &  34 \hspace{1pt}\text{mins}            & 7 \hspace{1pt}\text{mins}
                               & 0.8 \hspace{1pt}\text{s}\\
 \vspace{4pt}(2048,2080,3200)  &  \text{N/A}       & \text{N/A}
                               &   54 \hspace{1pt}\text{mins}           & 9 \hspace{1pt}\text{mins}
                               &  \text{N/A}                            & \text{N/A}
                               & 1.2 \hspace{1pt}\text{s}\\
 \vspace{1pt}\text{linear-$k$ functions}\hspace{1pt}(\text{nb.})&         &        &        &&&\\
 \vspace{0pt}(\hspace{4.5pt}128\hspace{0pt},\hspace{5pt}160\hspace{0pt},\hspace{5pt}240)
                               &    20 \hspace{1pt}\text{s}     & 2 \hspace{1pt}\text{s}
                               &    1 \hspace{1pt}\text{s}      & 0.1 \hspace{1pt}\text{s}
                               &    1.2 \hspace{1pt}\text{s}    & 0.5 \hspace{1pt}\text{s}
                               &    0.002 \hspace{1pt}\text{s}\\
 \vspace{0pt}(\hspace{4.5pt}256\hspace{0pt},\hspace{5pt}280\hspace{0pt},\hspace{5pt}400)
                               &    1\hspace{1pt}\text{min}    & 6 \hspace{1pt}\text{s}
                               &    4\hspace{1pt}\text{s}       & 0.4 \hspace{1pt}\text{s}
                               &    4.3\hspace{1pt}\text{s}     & 0.6 \hspace{1pt}\text{s}
                               &0.03 \hspace{1pt}\text{s}\\
 \vspace{0pt}(1024,1050,1600)  &    \text{N/A}    & \text{N/A}
                               &    6\hspace{1pt}\text{s}       & 0.8 \hspace{1pt}\text{s}
                               &    6.5\hspace{1pt}\text{s}     & 0.9 \hspace{1pt}\text{s}
                               &0.1 \hspace{1pt}\text{s}\\
 \vspace{0pt}(2048,2080,3200)  &    \text{N/A}    & \text{N/A}
                               &    20\hspace{1pt}\text{s}       & 1.6 \hspace{1pt}\text{s}
                               &    22\hspace{1pt}\text{s}     & 1.7 \hspace{1pt}\text{s}
                               &1.2 \hspace{1pt}\text{s}\\
\end{array}\]
 \vspace{12pt} 
 \tcaption{Performance of cryptovaluations on the three environments,
 where the term ``{\rm nb.}" stands for the encrypted computation without blindness
 for linear-$k$ functions, {\em i.e.}, addition, subtraction and string comparison.~\label{Fig-bks-EHE-laptop-HPC}}
\end{table}
 \vspace{0pt}

 Apropos of the, sectional, cryptovalaution governed by the 2nd code, likewise,
 the triplet $(k,w,n)$ codifies the three parameters of encrypted polynomial sets ${\cal P}_{n,w}({\cal U}_{cv,\hspace{1pt}q};\bm{z})$.
 To achieve the {\em blind} computation,
 the encrypted functions should be {\em indiscernible} during the computation.
 For this purpose,
 the runtimes of generating encrypted polynomials are adjusted to be close.
 The number of sections is in the range $n/2\leq e\leq 4n$ on the single-CPU
 and $n/8\leq e\leq n$ on the other two arenas.
 Symbols $T_{kg\text{-}sc}$, $T_{kg\text{-}mc}$ and $T_{kg\text{-}sg}$ record the longest task span of the polynomial generation among sections respectively on the single-CPU,
 multiple CPUs and single-node GPU,
 $T_{evl\text{-}sc}$, $T_{evl\text{-}mc}$ and $T_{evl\text{-}sg}$ the evaluation times,
 and $T_{de\text{-}sc}$, $T_{de\text{-}mc}$ and $T_{de\text{-}sg}$ the decoding times.
 The complete temporal course covers the time of data read and communication,
 conforming to the proportions the same as those of IME.
 As displayed in Table~\ref{Fig-bks-EHE-laptop-HPC},
 attributed to the enlargements of parallelism, memory and bandwidth on the multi-CPUs and single-node GPU,
 the generation of encrypted polynomial sets and the polynomial evaluation gain a performance raise with a factor of 10 to 20 times over the baseline of single-CPU.
 It is hard for a known HE to elevate efficiencies taking advantage of parallelism primarily
 restricted to the sequential nature of the recursive noise reduction.
 The deciphering times are comparable on the three computing playgrounds.
 The parameter triplet maximally reaches $(256,280,400)$ on the single-CPU,
 $(1536,1560,2400)$ on multiple CPUs and $(1024,1050,1600)$ on the single-node GPU.
 In a word,
 EHE receives the edge of encrypted computations of grand sizes
 far overtaking the limitation of subsisting HE.
 Notice that if the blindness is lifted from cryptovaluations of linear-$k$ functions,
 conducted in simpler encryptions with lower numbers of sections,
 the key-generation and encoding times reduce to a factor of ten or more and the plaintext size experiences a 1.5-fold expansion minimally.
 The effectiveness of the processing speed and memory usage will be additionally granted if
 replacing the 64-bit computing units with precise single-bit operations~\cite{2016-1-bitCom-Chmiel,2023-1-bitCom-YLi}.

 Obtained on relatively non-advanced machines, outcomes presented are preliminary and for temporary reference only.
 Sustained optimizations and continual enhancements of EHE software will align with the upgrades of computing hardwares.
 Readers are advised to consult the most up-to-date information provided on the software website~\cite{EHE-software-test-web}.
 As exhibited in the experiment, the larger the scale of the problem increases,
 the more pronounced the merit of EHE becomes over the existing HE.
 Explicitly, the advantageous puissance of EHE originates from
 practising {\em intrinsic parallelism} successively on the circuit segmentation, polynomial generation, polynomial evaluation and monomial calculation.
 Bordering on the foundational level of language exploited in machine code,
 invertible gates are transparently appropriate for manufacturing energy-economic systems~\cite{1961-Landauer,1991-Landauer,2010-ReverseGateDesign-Maha,2014-ReverseGateDesign-Jamal}.
 To substantially strengthen the potency of this framework,
 the suggestion is set forth on fabricating dedicated hardwares supporting the massive parallelism,
 the great amount of memory, the rapid data access and transfer,
 cores affording minimally necessary functions, and the accurate single-bit computation.
 Incorporating with the optimized software,
 a such dedicated hardware will be tailored to different computing environments
 and forged expectantly as a portable device through the miniaturization technology.
 This vision will considerably extend the scope of EHE to encompass a widespread spectrum of applications.

 \section{Conclusion}\label{secConclude}
  \renewcommand{\theequation}{\arabic{section}.\arabic{equation}}
\setcounter{equation}{0}\noindent
 In this article, the EHE framework is proposed,
 which merges concepts from two fields recently attracting great attention,
 {\em quantum computation} and {\em cryptography}.
 Significantly,
 {\em quantum gates} are introduced to EHE,
 substituting for non-invertible logic operations used in finite computations.
 Apart from exerting on quantum states conventionally,
 each quantum gate acts on {\em variables} to generate {\em polynomials}.
 This newfangled perspective capacitates the actualization of enciphering message and computation respectively through an encryption transformation
 formed in a product of quantum gates randomly chosen.
 Thanks to the succinct duality relation,
 a ciphertext is the evaluation of a polynomial set on an input plaintext,
 and then the outcome of encrypted computation is the evaluation of an encrypted polynomial set on the ciphertext.
 Disparate to prolix cryptograms of the two major post-quantum cryptosystems,
 the size of ciphertext hither is {\em compact}.
 The success of EHE is rooted in two essential traits of quantum gates, the {\em invertibility} and the {\em noncommutativity}.
 Rather than the approximated scheme of an existent HE,
 an {\em exact} encrypted computation is fulfilled in virtue of {\em invertible} gates.
 An {\em accurate} decryption is induced congruously,
 trouncing noisy deciphering of current encryptions.
 Stemming from the hardness of solving approved intractable problems
 and a {\em combinatorially} high complexity of reconstructing a {\em circuit} of {\em noncommuting} gates,
 an advanced level of security is reaped individually in the two encryptions.
 The {\em blindness} of homomorphic computations is further attained
 founded on the indiscernibility of encoded functions.
 Confronting quantum adversarial attacks,
 both encryptions exceed the standard security $2^{128}$ of {\em quantum resilience},
 and better surpass the suggested threshold of security $2^{1024}$ of {\em hyper quantum resilience}.
 Since each activated gate is {\em dimension-one preserving},
 EHE is straightforwardly realizable on classical computing environments without relying on quantum computers.
 The software test manifests that the {\em inherent parallelism},
 hierarchically wielded in EHE, is conductive to efficient and large-scale encrypted computations.

 The preservation of privacy and confidentiality has risen as a major concern in all aspects of modern life.
 With the advent of novel algorithms and computing technology,
 peculiar menaces are incurred to the protection of sensitive information.
 The EHE framework inaugurates a unique opportunity to acquire a full safeguard of secretiveness across different domains of academia and application zones.
 This framework illuminates an unexplored avenue of the research field of {\em noncommutative cryptography},
 which elucidates territories for future study.
 It forwardly offers preeminent methodologies
 to tackle intricate problems~\cite{1984-Learning-Valiant,1990-LearningDept3-J.H,2010-LearningCircuit-SY,2016-LearningCircuit-YS,2019-LearningDept3-KS}
 and inventive concepts to create effective tactics of cryptographic defenses.
 Meanwhile, the propounded software-hardware blueprints will initiate extraordinary usage throughout a multitude of real-world potential areas.
 In conclusion,
 EHE is perceived with promise that possesses a profound influence of furnishing prevailing guardianship in the landscape of cybersecurity.

 \vspace{20pt}
 \noindent{\bf Acknowledgements}\\
 \\
 Z.-Y.S. acknowledges National Center for High-Performance Computing and
 National Applied Research Laboratories, Taiwan, R.O.C.
 for the supports of constructing the framework Exact Homomorphic Encryption.
 M.-C.T. is grateful to National Center for High-Performance Computing in the support of developing the software.

\nonumsection{References}

\end{document}